\documentclass[conf]{new-aiaa}
\usepackage[utf8]{inputenc}
\usepackage{graphicx}
\usepackage{amsmath}
\usepackage[version=4]{mhchem}
\usepackage{siunitx}
\usepackage{longtable,tabularx}
\usepackage{subcaption}
\usepackage{mwe}
\usepackage[section]{placeins}
\usepackage{comment}
\usepackage{xcolor}

\usepackage[normalem]{ulem}
\usepackage{soul}

\title{Sparse flow reconstruction methods to reduce the costs of analyzing large unsteady datasets}

\author{ Spencer L. Stahl\footnote{AFRL Science and Technology Fellow} and Stuart I. Benton\footnote{Senior Research Aerospace Engineer, Aerospace Systems Directorate, AIAA Associate Fellow.\newline\newline Distribution Statement A: Approved for Public Release; Distribution is Unlimited. PA\# AFRL-2024-5787.}}
\affil{Air Force Research Laboratory, Wright-Patterson AFB, Ohio, 45433}

\begin{document}

\maketitle

\begin{abstract}


The cost of writing, transferring, and storing large amounts of data from unsteady simulations limits the accessibility of the entire solution, often leaving the majority of the flow under-sampled or not analyzed. 
For example, modeling the transient behavior of rare, but important, dynamic events requires three-dimensional snapshots written at high sampling rates, over a long duration. 
As such, the simulation time needed and large quantity of data produced, makes this a challenging problem for practical computational fluid dynamic (CFD) workflows, where memory resources are often limited and the writing penalty for modern GPU computing is much costlier.  
In this work, multiple sparse flow reconstruction (SFR) methods are developed to approximate a full unsteady solution by writing far fewer sparse measurements from the CFD solver, thus diminishing writing costs, data storage, and enabling greater sampling rates.
SFR is motivated by a large-eddy simulation (LES) example pursuing rare inlet distortion events, demonstrating that a down-sampling in full snapshots, supplemented by high-frequency sparse measurements, can substantially reduce writing time for a GPU solver and nearly eliminate the writing cost for a CPU solver.
In its simplest form, the "snapshot" SFR method is a single equation and can be further compressed with Proper Orthogonal Decomposition (POD-SFR) or its smaller and faster double POD-SFR variant.
A streaming SFR modification reconstructs snapshots more efficiently when local memory cannot store the entire solution. 
A sensitivity study evaluates the SFR scaling trade-off between sparse sampling rates and reconstruction accuracy, outlining best practices. 
To offset error of using random sparse measurements, the SFR approach exactly preserves dynamics in designated flow regions by additionally specifying sparse measurement locations, used here to capture the inlet distortion events. 
Distortion events are evaluated using the conditional space-time proper orthogonal decomposition (CST-POD) to pursue physical insights that characterize the upstream causality at full resolution.
A validation study of CST-POD modes confirms SFR effectiveness at retaining the event dynamics with substantial computational and memory savings.



\end{abstract}

\section{Introduction}


Modern computational fluid dynamics (CFD), combined with continual hardware advancements, has enabled high-fidelity simulations of ever increasing size and speed \cite{Slotnick2014_CFD2030}. 
Scale resolved simulations routinely calculate Terabytes of three-dimensional data for a single unsteady solution, of which only a fraction of that can feasibly be analyzed through significant down-sampling in space and time. 
The full potential of CFD in science and engineering is therefore fundamentally limited by the cost of data in terms of time, bandwidth, and memory allocated to writing, transferring, storing, and analyzing a full solution.
Furthermore, the current CFD computing architecture shift from central processing units (CPU) toward graphical processor units (GPU) \cite{Morris2023} comes with a greater data writing penalty, slowing down simulations and diminishing GPU benefits.
From a flow physics perspective, the pursuit of significant, but rarely occurring, flow events further challenges data costs by requiring the writing of long-duration datasets at high-sampling rates to capture sufficient transient dynamics that may only exist over a short time-scale. 
A relevant propulsion example used here, is found in inlets and diffusers, where extreme three-dimensional distortion events in the unsteady pressure field can surge and stall downstream engine components \cite{Triantafyllou2015}.
As such, modeling distortion is critical for predicting engine safety margins, but often lacks a complete understanding of the most dangerous outlier events due to strenuous data costs. 

An appealing solution to this overarching data challenge is compressed sensing \cite{Davenport2009,Manohar2018}, a term somewhat interchangeable with sparse sampling, or sparse compression.
These related ideas are adapted here as a method to reconstruct high-resolution, three-dimensional, flow snapshots at a high-frequency sampling rate, calculated from far fewer sparse measurements written by the CFD solver to increase its efficiency.
To address the underutilized potential of this method, we propose several new sparse flow reconstruction (SFR) methods and demonstrate best practices to improve computation speed, while balancing data compression and accuracy.

There is a profitable history of leveraging sparsity principles in data compression algorithms, most ubiquitously encountered in the digital formatting of audio, image, and video files \cite{Davenport2009}.
Compressed sensing was founded in such signal processing as a means to surpass the limits of the Nyquist–Shannon sampling theorem \cite{Nyquist,shannon}, with the goal of representing continuous signals from fewer discrete samples, only possible when data is made sparse in some basis, such as in the time, space, Fourier, or wavelet domain.
Building from these concepts, the potential to compress and reconstruct massive spatial-temporal systems from sparse measurements has spurred several developments across a range of topics \cite{Kim2024,Zhang2023,Manohar2018,Yang2016,Callaham2018}.
In fluid dynamics, interesting applications include: predicting the off-surface flow-field from limited surface-mounted sensors \cite{Bright2013,Willcox2006}, control based on optimization of sparse sensor locations \cite{Fukami2024}, and assimilating two data sources sampled at different rates \cite{He2024}.
Related, are the more complex machine learning techniques that use similar ideas of "super-resolution" to interpolate flow-fields from sparse points \cite{Fukami2024_single,Fukami2023,Zhong2023,Dubois2022}.

Although the aforementioned research demonstrates insightful use of sparsity, the underlying math can vary substantially with each application and objective, ranging from linear algebra to neural networks. 
Yet, widespread adoption of sparsity in CFD workflows is lacking, and has not been considered to improve the problem of writing penalties exacerbated by GPUs.
This could be attributed to the complexity of black-box algorithms that are difficult to integrate with a CFD solver, or demonstrations that are typically applied to low rank systems of basic configurations.
Here with SFR, we advocate for the straightforward, single-equation, interpolatory projection framework of \citet{kracht2023}, requiring only a full-state basis derived from data and sparse measurements.
The SFR method adapts the general framework of such Discrete Empirical Interpolation Methods (DEIM)  \cite{Chaturantabut2010,Farazmand2024} by choosing the full-state basis to be full-sized snapshots, down-sampled in time, while sparse measurements are simultaneously sampled at the highest rate desired for the reconstruction.
Several additional SFR modifications are introduced in Secn.~\ref{secn:method} to further improve data compression and memory performance to enable post-processing analyses when the dataset is far larger than available memory.


One challenge of implementing SFR (as well as other sparse sensing methods), is determining where to place sparse measurements to maximizes the reconstruction accuracy \cite{Brunton2016}.
A seminal finding from compressed sensing research is the use of random measurements to efficiently sample incoherent flow features that span the entire solution space \cite{Mousavi_2019,Krahmer2014}.
More advanced methods such as convex relaxation, non-convex, "greedy" algorithms, and more \cite{Crespo_2019,Farazmand2023}, seek the optimal number and location of sparse measurements, but may be burdened by iterative or complex math such as recent neural network methods \cite{Luo_2024,Cheng2024} that require tuning parameters.
For the most pragmatic SFR method, a liberal use of randomly placed sparse measurements is preferred due to its simplicity over optimized methods, which require recursive sampling updates from the CFD solver.
This brute-force approach is justified because even writing tens of thousands of random sparse signals to span the three-dimensional domain is still much cheaper than writing and storing full snapshots on the order of millions of degrees of freedom (DOFs) at that higher sampling rate. 
It will also be demonstrated that the main source of reconstruction error comes from temporal down-sampling of full snapshots, not from the spatial sampling of sparse measurements.
However, this random approach lacks optimality, and the number of sparse measurements to use is not known \textit{a priori}, thus best practices must be developed; a major objective of this work in Secn~\ref{secn:snapshot_results}. 
To compensate for any local error from random measurements, or global error from temporal sampling rates, the SFR approach here emphasizes the addition of user specified sparse measurement locations, where full accuracy is guaranteed in those regions \cite{Chaturantabut2010}; this is particularly important for studying rare fluid dynamics events in a practical manner.

Also of note are the more fundamental reduced order models (ROMs) that do not use sparsity, but do employ modal decomposition to construct a basis of modes that can reconstruct the unsteady flow more efficiently than the temporal snapshots \cite{Taira2017a_modal}.
Contemporary examples include, proper orthogonal decomposition (POD) \cite{Lumley1967structure,Sirovich1987,Berkooz1993}, spectral POD \cite{Towne2018_relation}, and dynamic mode decomposition (DMD) \cite{Schmid2010}. 
However, with these ROM approaches, the full snapshot solution must first be produced, then decomposed and compressed by truncating the basis of modes, and then finally uncompressed (reconstructed) for analysis.
Unlike the SFR philosophy, this only alleviates the data storage and transfer problems, but not the expense of writing the full snapshots in the first place.
Despite this, the math of modal decomposition overlaps with sparse sensing in regard to exploiting a data-driven basis of modes that better represents the dynamics~\cite{Manohar2018}. 
This has led to sparsity improving multiple modal decompositions, such as gappy POD \cite{Everson1995}, compressed sensing DMD \cite{Brunton2013,Bai2020}, sparsity promoting DMD \cite{Jovanovic2014}, resolvent analysis \cite{lopez_doriga_2024}, nonlinear models \cite{Brunton2015_sindy}, and methods for control~\cite{Kaiser2018,Herrmann2023}.
As will be explored in Secn.~\ref{secn:POD_perf}, a POD-based approach can easily be integrated with SFR algorithms, yielding an order of magnitude further compression when POD modes form a truncated basis for the interpolatory projection.

After a reduced order model has been obtained, another challenge is that the reconstructed solution may be Terabytes of data, far exceeding available local memory to uncompressed the entire solution at once. 
Therefore, a SFR streaming modification is desired, where individual snapshots can be reconstructed one at a time, quickly for real-time visualization. 
The SFR streaming algorithm presented in Secn.~\ref{secn:method} is considered offline, where a smaller interpolation operator and the basis acquired from available snapshots is stored locally, with snapshots uncompressed and parsed when needed for analysis.  
Online streaming methods have been developed for modal decompositions like POD \cite{Li2022}, and could be used to iteratively update the POD-SFR basis as the CFD solver generates new snapshots, although this method is not implemented here.
However such streaming and parallelization algorithms reflect the fact that decompositions in general are expensive in terms of memory requirements and calculation time.
For the POD-SFR method, this inefficiency is resolved by a "double POD-SFR" variant, where sparse measurements are instead decomposed, followed by reconstruction of a full sized POD modes, then reconstruction of unsteady snapshots.
This faster approach can be combined with the streaming algorithm to address all memory bottlenecks for CFD post-processing.

Demonstration of SFR on realistic, high-Reynolds number, complex problems is necessary to asses the validity of the method for practical CFD use.
Here we test SFR on LES data of an inlet-isolator-diffuser, featuring transonic flow (Mach 1.7) further complicated by three-dimensional corners and intricate instrumentation geometry \cite{stahl_sci_2024,stahl_sci_2025}. 
This highly turbulent flow ($Re=2\times10^6$, based on throat height) is characterized by multiple shock-boundary-layer interactions, flow separation, and ultimately downstream distortion.
Of engineering concern is how these phenomena contribute to total pressure losses and unsteady distortion patterns at the diffuser exit, which propagates instabilities to jet engine components, severely affecting operability and performance.  
While time-averaged and second-order statistics of these metrics can be calculated without incurring the costs of writing the full unsteady solution, fully capturing the unsteady dynamics of rare and extreme pressure events does.
In particular, the difficulty lies within the large quantity of data needed at high sampling rates over long time periods to gather a sufficient number of events; a problem well-suited for SFR.
After this data is obtained using SFR, the rare events are studied in Secn.~\ref{secn:CPOD} using the Conditional space-time proper orthogonal decomposition (CST-POD) \cite{stahl_2023_CPOD_jcp,Schmidt2019}, a valuable diagnostic tool for identifying and delineating underlying event dynamics across a set of modes ranked by energy.  
In addition to analysing global error, CST-POD modes calculated using SFR are validated to determine how accurately distortion events are captured with varying levels of sparsity.
This paper culminates in Secn.~\ref{secn:conclusions} with a summary of SFR best-practices to balance accuracy and data compression. 

\section{SFR methodology}\label{secn:method}

\subsection{General SFR Formulation}

\begin{figure}
\centering
\includegraphics[width=1\textwidth]{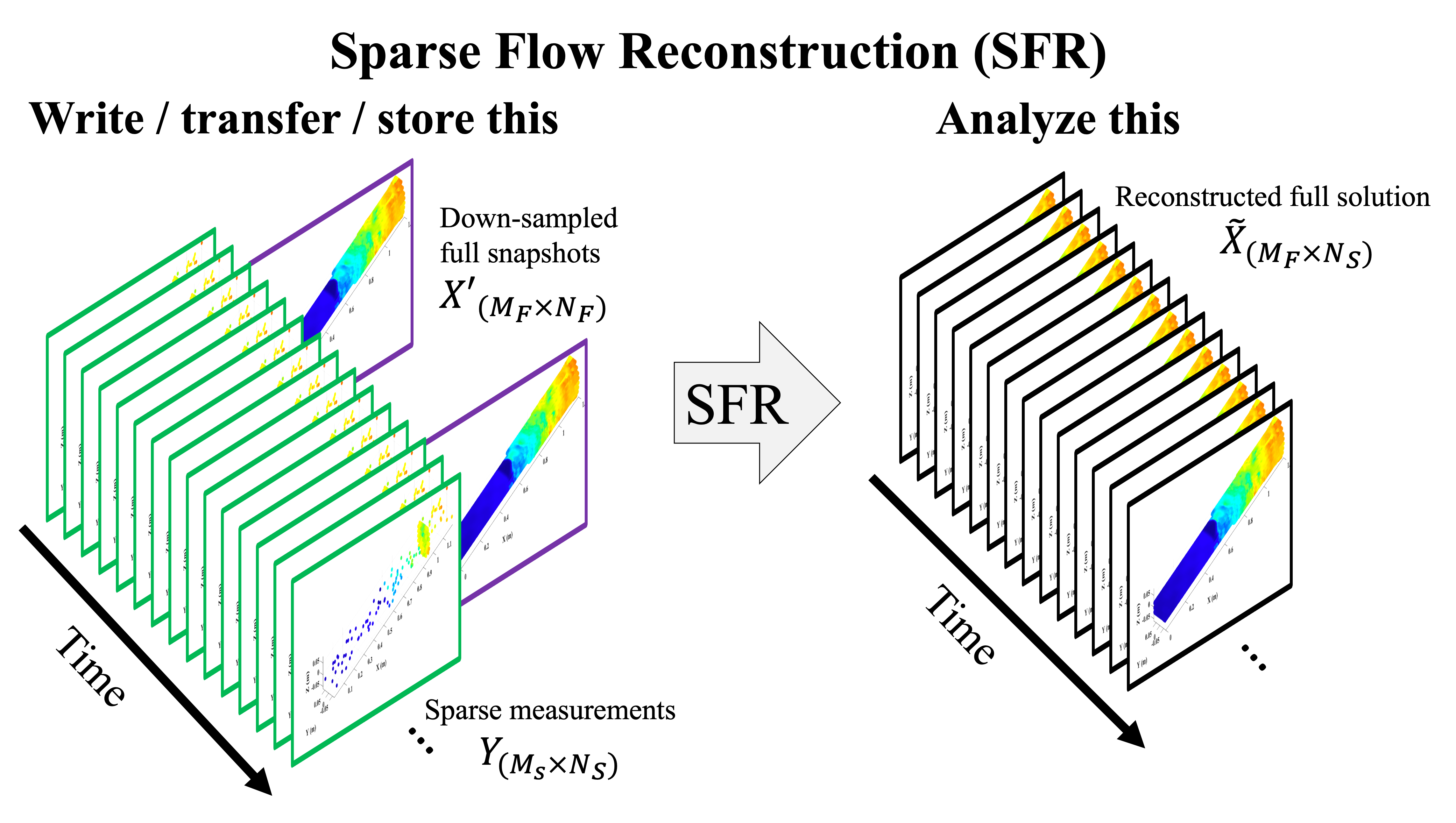}
\caption{ The Sparse Flow Reconstruction (SFR) method writes, transfers, and stores two smaller sparse datasets: the temporally down-sampled full snapshots, $\textbf{X}'$,  and high-frequency sparse measurements, $\textbf{Y}$. The SFR method reconstructs the full snapshot solution at the higher sampling rate of the sparse measurements, $\tilde{\textbf{X}}$. This framework improves CFD writing and post-processing efficiency.} 
\label{fig:SFR_intro}
\end{figure}

The SFR formulation is designed to reconstruct long-duration, high-sampling rate, discretized flow data in the form of a standard snapshot matrix, $\mathit{\mathbf{X}}_{M_F \times N_S}$, where $M_F$ is the degrees of freedom throughout the full-sized three-dimensional domain.
A total of $N_S$ snapshots are arranged as the columns of \textbf{X} and are taken at the highest necessary sampling rate $f_S$.
We assume this snapshot matrix is massive, and incurs large penalties in writing, transfer, and storage, which we seek to avoid. 
The SFR method circumvents this in the manner depicted in Fig.~\ref{fig:SFR_intro}, where the full solution is instead approximated, $\mathit{\mathbf{\tilde{X}}}_{M_F \times N_S}$, by combining two smaller datasets, sparse in space and time. 
These include: 1) the temporally down-sampled full spatial domain snapshots $\mathit{\mathbf{X'}}_{M_F \times N_F}$, taken at a lower sampling frequency, $f_F$,  where $N_F < N_S$,
and 2) the sparse measurements, $\mathit{\mathbf{Y}}_{M_S \times N_S}$, taken at far fewer spatial locations ($M_S << M_F$) but at the higher sampling rate of the full solution $f_S$, or in practice, at whatever temporal resolution is desired. 
Details of the sparse measurement locations and their random sampling are discussed later, along with additional opportunities for POD compression and algorithm improvements. 
For now, the fundamental idea is that by cleverly combining the two sparse datasets, the complete solution can be reconstructed with minimal error, forgoing the cost of writing all snapshots, as this data can be uncompressed on the fly for post-processing analysis.


The most general formulation to calculate the SFR solution, $\tilde{\textbf{X}}$, follows the framework outlined in \citet{kracht2023}, using the mathematical theory of DEIM \cite{Chaturantabut2010}, which is interpreted as an interpolatory projection of sparse measurements $\textbf{Y}$, onto a general set of basis vectors $\textbf{B}$, that represents and spans the full "truth" snapshot matrix, $\textbf{X}$.
The fundamental SFR equation to approximate the solution, $\tilde{\textbf{X}},$ for a generic basis matrix, $\textbf{B}$, is: 
\begin{equation}\label{eqn:SFR_gen}
    \mathbf{\tilde{X}=B(P^TB)^{-1}Y},
\end{equation}
where $\mathbf{P}$ is the sparse identity matrix that identifies the indices of the sparse measurements, arranged within the larger set of full snapshot indices.
Sampling interval gaps between the two datasets, $\textbf{Y}$ and $\textbf{X}'$, are interpolated beyond the Nyquist limit because of the high-frequency sparse measurements threaded into the projection of the basis $\textbf{B}$.

The most exact basis that spans the truth solution are the high-frequency full snapshots themselves (\textbf{X}), however the purpose of SFR is to avoid writing this much data.
Instead, the basis is narrowed to the temporally down-sampled matrix, $\textbf{X}'$, which still encapsulates relevant dynamics and modes.
Of the two input datasets needed for the SFR, sparse measurements, $\textbf{Y}$, are explicit in Eqn.~\ref{eqn:SFR_gen}, while the under-sampled full solution, $\textbf{X}'$, is used to construct the basis, $\textbf{B}$, which can take many forms \cite{kracht2023}.
The SFR method outlines two options for choosing $\textbf{B}$, comparing the advantages and disadvantages of implementing each throughout the paper:  
\begin{enumerate}
    \item The first and simplest SFR method, directly uses the down-sampled full snapshots as the basis, $\textbf{B}=\textbf{X}'$, and is referred to as \textit{snapshot SFR}. 
    \item The second ROM-based approach, implicitly uses the down-sampled snapshots to derive a POD basis, and is referred to as \textit{POD-SFR}. For this, the singular valued decomposition is employed, $\textbf{X}'=\textbf{U}\boldsymbol{\Sigma} \textbf{V}^T$, to acquire the POD modes (vectors of $\textbf{U}$). Functionally, a SFR basis containing a smaller truncated set of $r$ modes, $\textbf{B}=\textbf{U}_r$, provides additional and adjustable data compression. Including all POD modes ($r=N_F$) produces the exact same approximation as the snapshot SFR method, while truncating the number ($r<N_F$) reduces the size of \textbf{B}, for greater compression at the expense of error.
\end{enumerate}

\subsection{Snapshot SFR method}\label{secn:snap_SFR_method}

\begin{figure}
\centering
\includegraphics[width=1\textwidth]{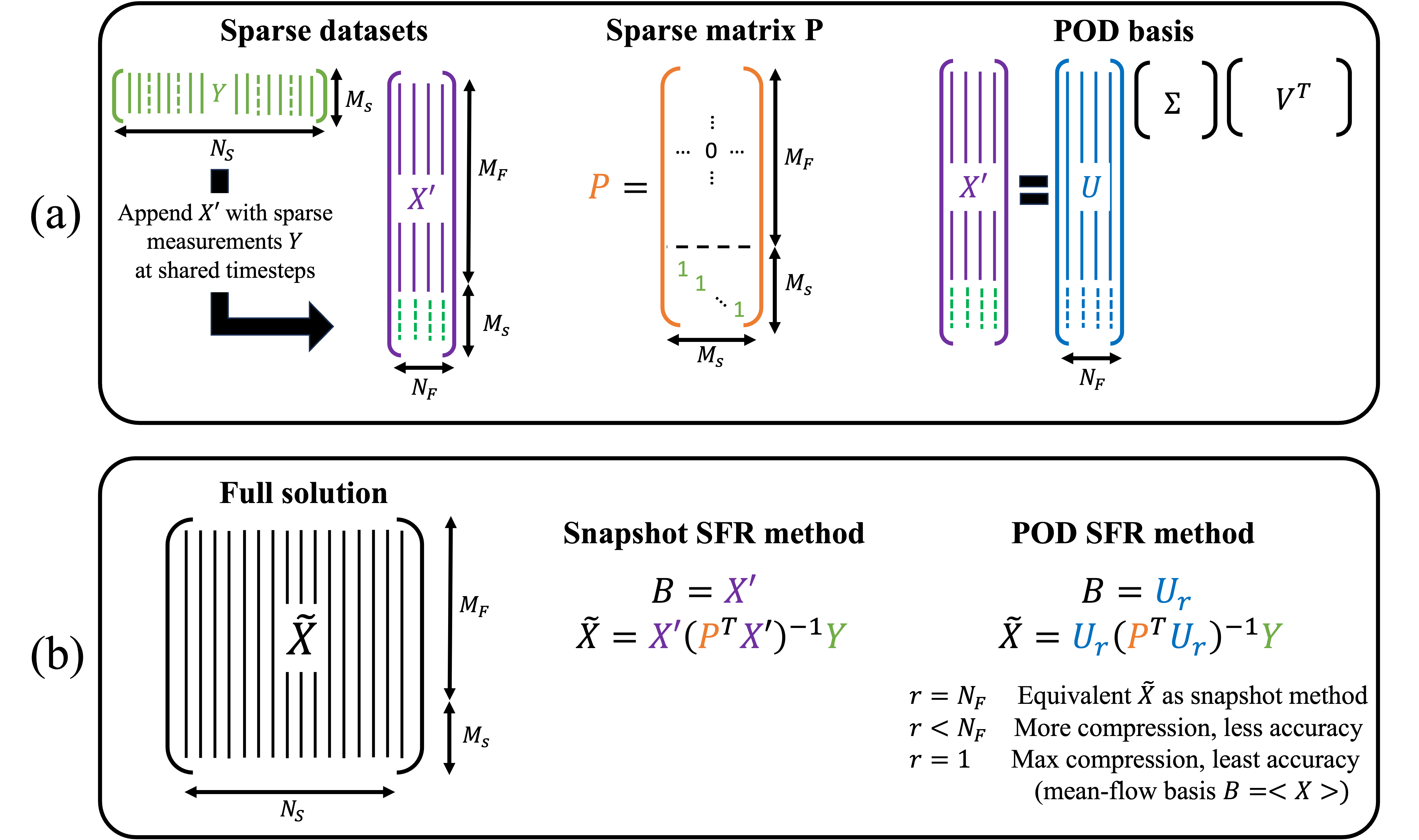}
\caption{(a) SFR data matrix structures and the (b) full solution reconstruction from the snapshot and POD-SFR methods. }
\label{fig:SFR_math}
\end{figure}

The basic features of the snapshot SFR method are discussed first. 
For reference, the input data matrices and equations are visualized in Fig.~\ref{fig:SFR_math}. 
For the snapshot SFR formula, the basis is, $\textbf{B}=\textbf{X}'$, and the full solution is therefore approximated by, 
\begin{equation}\label{eqn:SFR}
    \mathbf{\tilde{X}=X'(P^TX')^{-1}Y}.
\end{equation}
This is a straightforward approach and only requires a single equation to solve; in Secn.~\ref{secn:POD_SFR_method}, a streaming SFR modification breaks this calculation into multiple steps for more efficient memory use.
Note that in this work, the unstructured sparse measurement locations are additional to the full snapshot DOFs, as opposed to being a subset of a full snapshot; this requires appending shared time indices of $\textbf{Y}$ to the last rows of $\textbf{X}'$, as shown in Fig.~\ref{fig:SFR_math}(a).
Alternatively, if the sparse measurements are a subset of the full domain, the populated indices of $\textbf{P}$ should change accordingly to ensure all matrices are of consistent size for the SFR equations in Fig.~\ref{fig:SFR_math}(b). 
This implies that $\textbf{X}'$ and $\textbf{Y}$ should be sampled at common time-step intervals.
Typically this is done at integer multiple sampling rates, $f_S/f_F$, dubbed the \textit{interval} sampling approach which is interpreted as an interpolation problem.
In the \textit{consecutive} sampling approach, full and sparse samples are simultaneously sampled over an initial training period with $f_S=f_F$, followed by the remaining duration where only sparse measurements are written, forming a SFR extrapolation problem; both methods are compared in Secn.~\ref{secn:proxy_Error}.

Regarding the data reduction performance of SFR, a compression ratio, $C$, is defined by the size of the full dataset approximated, relative to the combined size of the two smaller input datasets,

\begin{equation}
    C=\frac{M_F\times N_S}{M_F\times N_F + M_S\times N_S},
\end{equation}
or rearranged as a function of the two SFR down-sampling parameters, $N_F/N_S$ and $M_S/M_F$, the compression is,
\begin{equation}
    C=\frac{1}{(N_F/N_S) + (M_S/M_F)}.
\end{equation} 
Since the size of the full snapshots matrix, $\textbf{X}'$, is usually much larger than the sparse measurements ($M_F \times N_F >> M_S\times N_S$), the compression rate is primarily driven by the down-sampling of the full snapshots in time, $N_F/N_S$.
Additional compression can be improved from a truncated POD basis, as discussed in Secn.~\ref{secn:POD_SFR_method}.

Alluding to the accuracy of the snapshot SFR approach, error is also driven by the number of full snapshots used, $N_F/N_S$, or equivalently for the interval sampling approach, their disparity in sampling frequencies, $f_F/f_S$.
Intuitively, the less snapshots and the longer duration between snapshots in $\textbf{X}'$, the less accurate the reconstruction will be, but with improved compression and writing costs. 
The number of sparse measurements, $M_S$, and their density relative to the full domain also contributes to reconstruction error.
Here, the practice of randomly sampling points throughout the three-dimensional domain sufficiently represents the dynamic system, improving as more random locations are added.
To compensate for spatial error in the random scattering of sparse probes, we importantly note the option to improve local accuracy by increasing the density of sparse measurements in regions of interest, as exemplified for the distortion problem in Secn.~\ref{secn:local_acc}.
Even a minimal contribution of specified high-frequency measurements included in $\textbf{Y}$, say at logical 2D slices of the three-dimensional domain, will perfectly preserve accuracy by overwriting the reconstructed solution at these location with the input measurements.
This technique is recommended since additionally specified sparse measurements are still few relative to the full snapshot size.

\subsection{POD-SFR method}\label{secn:POD_SFR_method}

A more efficient basis is exploited in the POD-SFR method, $\textbf{B}=\textbf{U}_r$, to further increase compression. 
The first step in the post-processing algorithm is the singular value decomposition, used to calculate POD modes, $\textbf{U}$, from down-sampled full snapshots, 
\begin{equation}\label{eqn:POD}
    \textbf{X}'= \textbf{U} \mathbf{\Sigma} \textbf{V}^T.
\end{equation}
The POD-SFR approach is possible because the snapshot matrices, $\textbf{X}, \tilde{\textbf{X}},$ and $\textbf{X}'$ all converge to the same POD modes as $N_F \rightarrow N_S$.
In practice, the POD basis from $\textbf{X}'$ will likely not be converged, and is intentionally truncated to size $r \leq N_F$ for additional compression.
That is, the last $N_F-r$ columns of $\textbf{U}$ are removed and the approximated basis becomes,
\begin{equation}
    \textbf{B}=\textbf{U}_r.
\end{equation}
The compressed POD-SFR reconstruction is then, 
\begin{equation}\label{eqn:SFR}
    \mathbf{\tilde{X}=U_r(P^TU_r)^{-1}Y}.
\end{equation}

When all POD modes are used ($r=N_F$), the basis spanned is equivalent to the snapshot SFR method and the results are the same; unless truncated to $r<N_F$, POD-SFR should be avoided since there is no benefit to the extra SVD step.
Again, the appeal of the POD-SFR approach, is in choosing a subset of mode for additional compression at the cost of accuracy.
For a low rank system, error may be insignificant if a few modes can represent the flow well.
Ideally, there is an optimal number of modes $r$ to keep.
In section~\ref{secn:POD_perf}, multiple orders of magnitude of compression are tested. 
Interestingly, choosing the extreme compression limit of the first POD mode, $r=1$, is equivalent to choosing the mean-flow as the single-vector basis, $\textbf{B}=\textbf{U}_1= <\textbf{X}'>$.
From the perspective of maximizing compression, the exact mean-flow can be used instead, $\textbf{B}=<\textbf{X}>$.
Advantageously, this is a byproduct of nearly all simulations, and would entirely eliminate the cost of writing unsteady snapshots and the SVD step, giving instantaneous access to the full unsteady snapshot data.
Of course, there is significant error associated with resolving the turbulent structures with only the mean-flow and a few sparse points and this extreme limit is not recommended.

Finally, note that a major problem with the POD-SFR method is that the SVD calculation may actually be too large to calculate from the full snapshots, $\textbf{X}'$, in terms of local memory available.
A solution is presented next with the \textit{double POD-SFR} method.

\subsection{SFR memory improvement options}\label{secn:mem_improvements}

\begin{figure}
\centering
\includegraphics[width=1\textwidth]{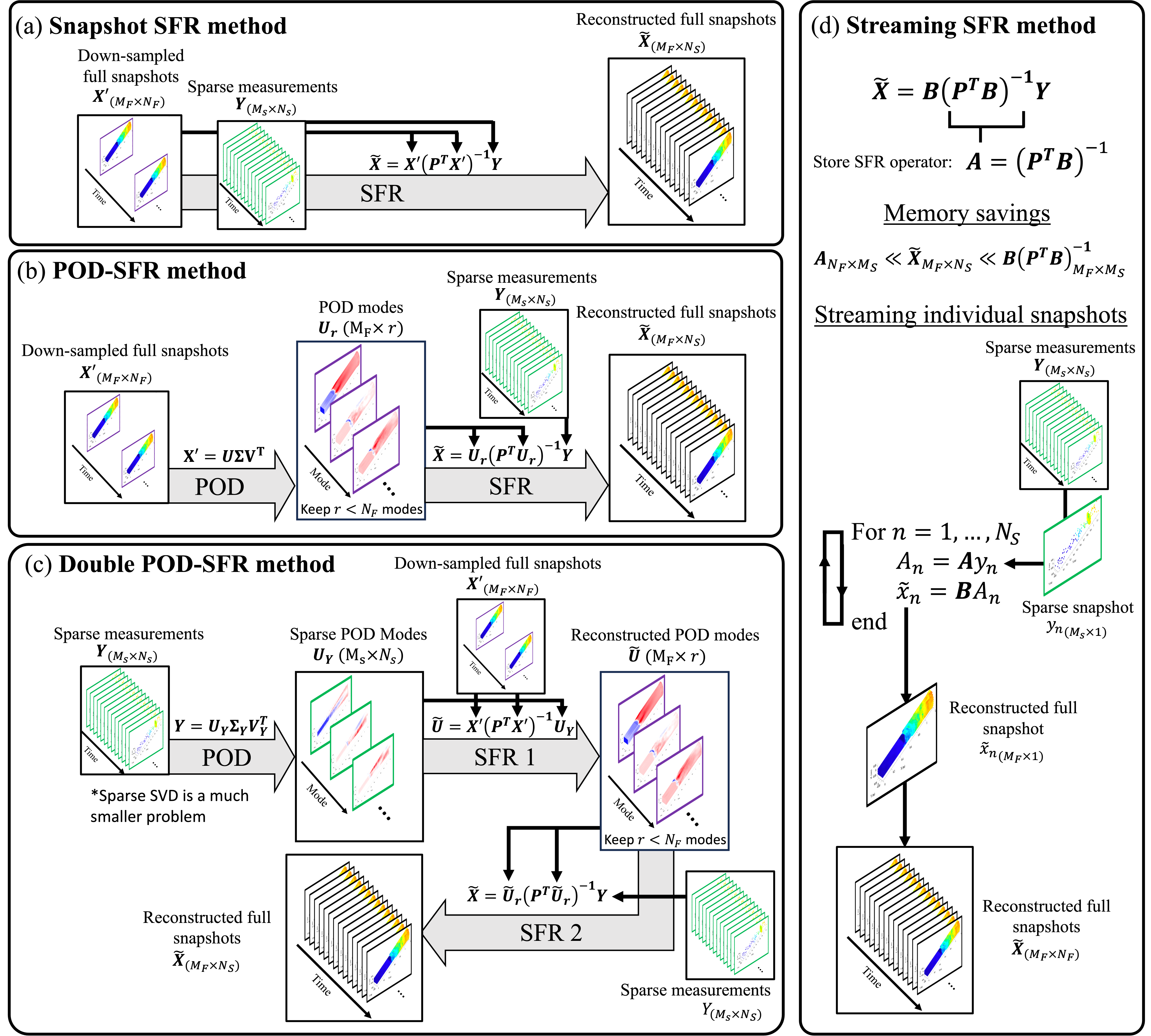}
\caption{(a) Snapshot SFR method and (b) POD-SFR method, where the POD basis is calculated from a SVD of all available full snapshots. (c) The double POD-SFR method more efficiently calculates the SVD on the smaller set of sparse measurements, followed by the first SFR to reconstruct a subset of $r$ full sized POD modes, forming the basis for the second SFR to reconstruct the full sized snapshots. (d) The streaming-SFR method can be applied to any SFR calculation to improve memory efficiency. }
\label{fig:ROM_SFR_method}
\end{figure}

A practical limitation of any data reduction method is the memory available in local RAM for post-processing analysis.
Specific to the POD-SFR method, calculation of the modal basis via the singular value decomposition of Eqn.~\ref{eqn:POD} may not be possible due to limited memory availability.
Even loading or storing the final SFR approximated dataset, $\tilde{\textbf{X}}$, may be impossible due to its large size.
Therefore, additional SFR algorithm modifications are outlined to address these problems, and visualized in the flowcharts of Fig.~\ref{fig:ROM_SFR_method}, with the basic (a) snapshot and (b) POD-SFR methods shown for reference.
A \textit{double POD-SFR} method is presented in Fig.~\ref{fig:ROM_SFR_method}(c), that drastically reduces the cost of calculating the POD basis, by instead calculating smaller POD modes from sparse measurements, followed by two calculations to reconstruct the full snapshots. 
Fig.~\ref{fig:ROM_SFR_method}(d), illustrates the \textit{streaming SFR} modification, which separates the snapshot or POD-SFR calculation into multiple steps, storing a smaller SFR operator in local memory, where individual snapshots can be then reconstructed "on the fly", eliminating the bottleneck of producing $\tilde{\textbf{X}}$ in a single shot when its size exceeds available memory. 

\subsubsection{Double POD-SFR method}

The modal decomposition required for the  POD-SFR compression method is expensive in terms of increased compute time and memory costs.
A more efficient double POD-SFR (Fig.~\ref{fig:ROM_SFR_method}(c)) leverages the sparse measurements to estimate the SFR basis.
First the POD modes of the sparse measurements, $\mathbf{U_Y}$, are calculated via the SVD,
\begin{equation}
     \mathbf{Y=U_Y \Sigma_Y V_Y^T}.
\end{equation}
This is a much smaller, and therefore much faster, singular value decomposition that circumvents the RAM requirements of obtaining the full-sized POD basis ($\textbf{B}=\textbf{U}$).
Next in the first SFR step, these full-sized modes are approximated and reconstructed, $\tilde{\textbf{U}}$, using an analogous SFR equation that substitutes the sparse time measurements, $\textbf{Y}$, with the sparse POD modes, $\textbf{U}_Y$, while using the down-sampled full snapshots as the basis $\textbf{B}=\textbf{X}'$,
\begin{equation}
        \mathbf{\tilde{U}=X'(P^TX')^{-1}U_Y}.
\end{equation}
Note the sparse reconstructed POD modes produced, $\tilde{{U}}$, may differ from truth POD modes, ${U}$, in terms of error matching their spatial patterns due to the sparse approximation.
Despite any discrepancy of particular modes at this intermediate step, the collective linear combination of all reconstructed modes, $\tilde{\textbf{U}}$, in the second SFR step, provides a sufficient basis $\textbf{B}=\tilde{\textbf{U}}$, that as will be shown, can exactly reproduce the performance of the snapshot SFR method if all POD modes are used.
Similar to the POD-SFR method, a trade-off in accuracy for compression is made when the basis is truncated to $r < N_F$ POD modes. 
To recover the final reconstructed snapshots, $\tilde{\textbf{X}}$, the truncated set of approximated POD modes, $\tilde{\textbf{U}}$, is combined with the temporal sparse measurements, \textbf{Y}, in the second SFR step,
\begin{equation}
    \mathbf{\tilde{X}=\tilde{U}_r(P^T\tilde{U}_r)^{-1}Y}.
\end{equation}
Note the SFR calculations here can also be replaced with the streaming method, shown next, used to calculate one mode or snapshot at a time.
This is particularly beneficial in the second SFR step since only a truncated set of POD modes $r$ are needed.



\subsubsection{Streaming SFR}

The ultimate SFR local memory requirement is determined by the size of the final SFR matrix, $\tilde{\textbf{X}}$.
This limitation is addressed by streaming each snapshot one at a time, rather than calculating the entire matrix as in Eqn.~\ref{eqn:SFR_gen}. 
For this, the SFR calculation is broken into three intermediate calculations, as outlined in Fig.~\ref{fig:ROM_SFR_method}(d).
First the interpolation operator is partially calculated,
\begin{equation}
    \mathbf{A=(P^TB)^{-1}}.
\end{equation}
This $\textbf{A}$ matrix is size $N_F \times M_S$ and is stored in memory. 
For a size comparison, $\textbf{A}$ is significantly smaller than both the final solution $\tilde{\textbf{X}}$ or the full sized operator $\mathbf{B(P^T B)}^{-1}$.
The next two calculations are performed in an offline streaming manner inside a loop, one snapshot, $n$, at a time.
First the instantaneous vector operator, $a_n$, is calculated from sparse measurements at that timestep,
\begin{equation}
    a_n=\mathbf{A} y_n.
\end{equation}
The $a_n$ vector comprises the coefficients of each basis vector in $\textbf{B}$ that are active at that time-instance, multiplied to obtain the reconstructed full snapshot $\tilde{x}_n$
\begin{equation}
    \tilde{x}_n=\mathbf{B}a_n.
\end{equation}

Calculating individual snapshots in this manner is much faster than calculating the entire solution $\mathbf{\tilde{X}}$ at once, and also opens the possibility of parallelization.
Snapshots can be discarded after use to minimize memory allocation, making this very efficient for visualizing large unsteady flow-fields in a streaming manner.
Alternatively, the individual reconstructed snapshots can be collected when needed for aggregate analysis.
The memory of the streaming SFR approach is now limited by the size of the basis stored $\textbf{B}$, and comes with no accuracy penalty.
This streaming approach can also be used with either the snapshot or POD-SFR methods, and is implemented for both in this work.


\section{Motivation: The cost of writing large unsteady data}\label{secn:write_test}
The need for SFR is motivated by the inefficient cost of writing unsteady snapshots in CFD workflows, and the philosophy of reconstructing large datasets while requiring less snapshots to be written. 
To preface, writing inefficiencies vary for every CFD solver, flow physics, mesh, hardware, ect. used. 
However, the trends and limitations presented next are valid for any situation where large unsteady datasets are to be written, and are most relevant for LES- or DNS-scale simulations. 
Furthermore, SFR is implemented as a post-processing technique and can therefore be used with any CFD solver.
Typical writing costs are instantiated here using the wall-modeled LES solver, CharLES \cite{Lakebrink2019_CFD_compare}, testing both CPU and GPU write penalties.
CharLES models the unsteady, three-dimensional, compressible Navier-Stokes equations using the Vreman sub-grid scale and equilibrium wall-model \cite{Vreman2004}.
A second-order spatial discretization scheme is used, with an explicit, three stage, Runge-Kutta scheme for time integration \cite{Gottlieb2001}.
Writing performance is tested on the Mach 1.7 inlet-isolator-diffuser problem of \citet{stahl_sci_2025} using the same geometry, boundary conditions, and mesh.
Details of the flow are discussed in Secn.~\ref{secn:snapshot_results}.

CFD writing costs are exemplified with the following computational experiment, where performance is determined by the number of time-step iterations calculated (or flow seconds simulated) in one hour of wall-clock compute time.
The writing penalty compares the number of time-steps simulated while writing unsteady three-dimensional snapshots, to a case where no snapshots are written, which represents the baseline maximum performance.
In all cases, the LES is calculated on a fixed mesh size of approximately 15 million (m) control volumes, representing a moderate sized CFD job.
Writing penalties are evaluated for different full-sized snapshots encompassing all five flow variables, and testing a range of sizes from downsampled to oversampled: $M_F=2m$, $9m$, and $18m$. 
For a relatively fair comparison of CPU and GPU resources used, calculations were performed with 6 compute nodes on the AFRL HPC system Raider, using all available 768 CPU cores (AMD 7713 Milan) or 24 GPUs (NVIDIA A100 SXM 4).
This ensures both setups deliver comparable computational throughput relative to their architectures and that simulation costs are equivalent in terms of HPC nodes reserved.

A physically ideal, but highly expensive sampling rate of the solution is tested, writing full snapshots, $\textbf{X}$, every $50$ iterations ($f=40~kHz$).
The full write performances are compared with the SFR snapshot method which reconstructs the full dataset, $\tilde{\textbf{X}}$, by writing fewer down-sampled snapshots $\textbf{X}'$.
In this SFR example, a $C=20$ data compression is targeted by lowering the full snapshot sampling rate to every $1000$ iterations ($f_F=2~kHz$), supplemented by writing fewer sparse measurements ($M_S=62,000$), in the data matrix $\textbf{Y}$, maintained at the higher $f_S=40kHz$ sampling rate.
For reference, simulating a single second of flow at these sampling rates would generate and require 1.16 TB of data storage for the 18m snapshot size, which is reduced to only 96.7 GB for the SFR method.
Appropriate SFR compression levels are informed by the error sensitivity study discussed later in Secn.~\ref{secn:snapshot_results}, but for now, demonstrate the potential of this method for improving CFD efficiency.

\begin{figure}
\centering
\includegraphics[width=1\textwidth]{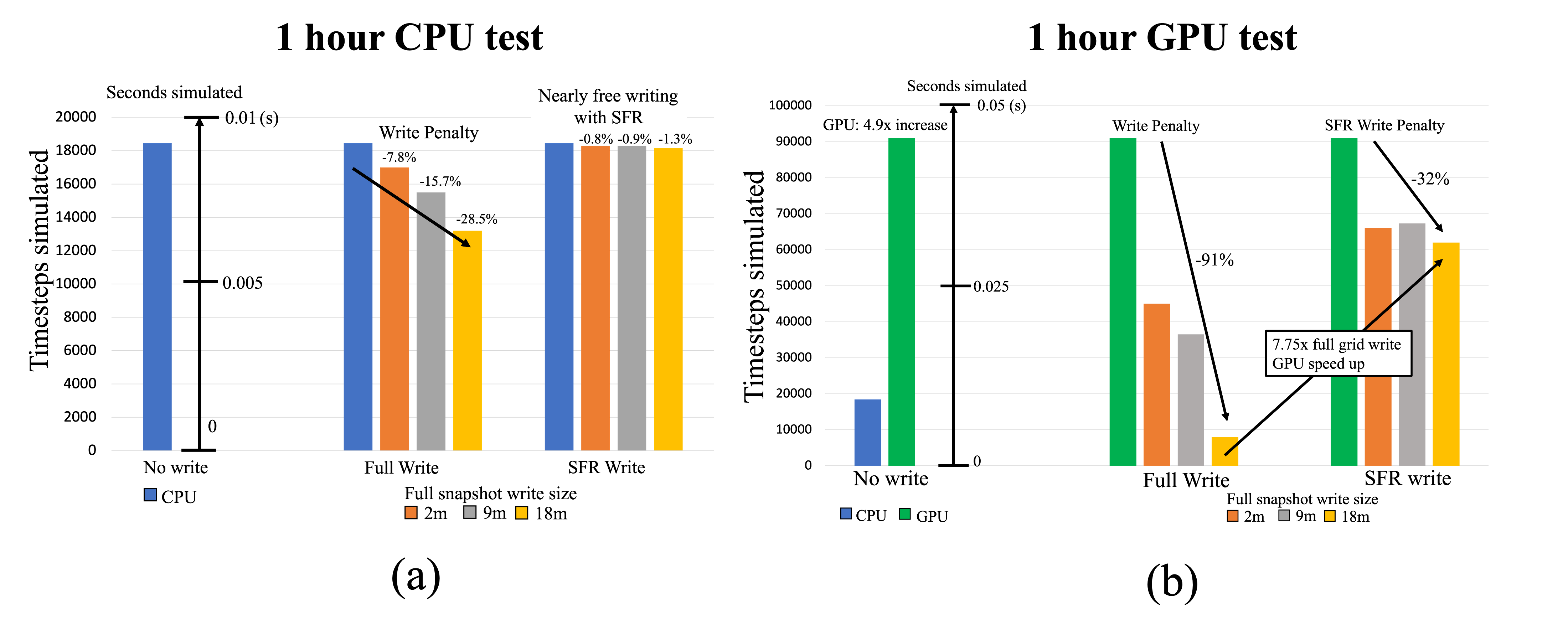}
\caption{The cost of writing large CFD datasets for (a) CPUs and (b) GPUs, compared to the SFR sparse measurements. All cases compare the number of iterations and seconds computed in 1 hour on a 15m grid. The full snapshot sampling rate of $40kHz$ is reduced to $2kHz$ for the SFR sampling rate. }
\label{fig:write_cost}
\end{figure}

Test results are presented in Fig.~\ref{fig:write_cost} for (a) CPUs and (b) GPUs, along with the improved SFR performance.
The CPU results are first discussed.
Without writing any full snapshots, one hour of compute time calculates 18k iterations - or approximately 0.009 seconds of flow time. 
The iterations calculated while writing full domain snapshots shows a sizable penalty, with a $28\%$ reduction in performance for the largest 18m full snapshots. 
On the other hand, the SFR down-sampling performed excellent with only a $1\%$ writing penalty, nearly matching the no write case. 
This means the small amount of data written to reconstruct the $1.16$ TB solution in post-processing is effectively free in terms of slowing down the simulation

A stronger case for SFR is made with the GPU results, where writing penalties are much costlier.
The baseline no-write GPU test in Fig.~\ref{fig:write_cost}(b) displays the enormous potential for this technology, increasing the iterations simulated in one hour to nearly $5$ times the CPU performance (0.045 flow seconds).
Unfortunately, even when down-sampled three-dimensional unsteady snapshots are written, the penalty is severe, more than halving the time-step iterations.
The largest 18m full snapshots have a devastating $91 \%$ writing cost, making this case less effective than the CPU solver.
GPU performance is recouped by the SFR method, which improves the writing penalty to approximately $30\%$.
For the largest snapshot size, this is a $7.5\times$ improvement, making the GPU solver viable again. 
While these specific results are unique to the inlet distortion problem and LES implementation, similar trends for other CPU and GPU based solvers likely follow for a broad range of problems.


\section{Snapshot SFR Performance}\label{secn:snapshot_results}

\subsection{Overview}

Characterizing the performance of the SFR method is critical for evaluating data compression and accuracy, as well as to outline best practices.
For this, the simpler snapshot SFR method is evaluated in this section to demonstrate fundamental trends and principles before examining the additional compression of the  POD-SFR approach in Sec.~\ref{secn:POD_perf}.
Accuracy is measured in terms of local and global error, in aggregate and as a function of time; later in Secn.~\ref{secn:CPOD}, the preservation of rare event dynamics will also be considered. 
First, the approach of placing sparse points in regions of interest is demonstrated as a means to guarantee local accuracy.  This is followed by a global accuracy sensitivity study that tests the progressive compression of sparse sampling parameters in time ($N_F/N_S)$ and space ($M_S/M_F)$. 
Other topics discussed include using sparse measurements as a proxy for estimating error when no truth solution is available, and the temporal error behavior of the full snapshot interval or consecutive sampling approaches.

The LES from the previous writing test in Secn.~\ref{secn:write_test} is used again, but now run to completion with variable sparsity parameters.
As a reference to the inlet problem, an instantaneous snapshot of the solution is presented in Fig.~\ref{fig:inlet_intro}(a), featuring the incoming Mach 1.7 flow ($Re=2\times10^6$) in the rectangular isolator section of an inlet, expanding to a circular diffuser, ending with the modeled rake instrumentation section.
Several aerodynamic phenomenon complicate the flow, including the oscillating shock train comprised of several shock boundary layer interactions (SBLI), and associated flow separation with regions of reversed flow.
Of particular interest, is the distortion measured at the aerodynamic interface plane (AIP), the circular cross-section where turbulence is characterized for inlet-engine compatibility. 
For this application, it is important that rare, high-pressure, distortion events are captured, thus motivating more sparse measurements at the AIP to accurately reconstruct all events.


\subsection{Local accuracy behavior}\label{secn:local_acc}

\begin{figure}
\centering
\includegraphics[width=1\textwidth]{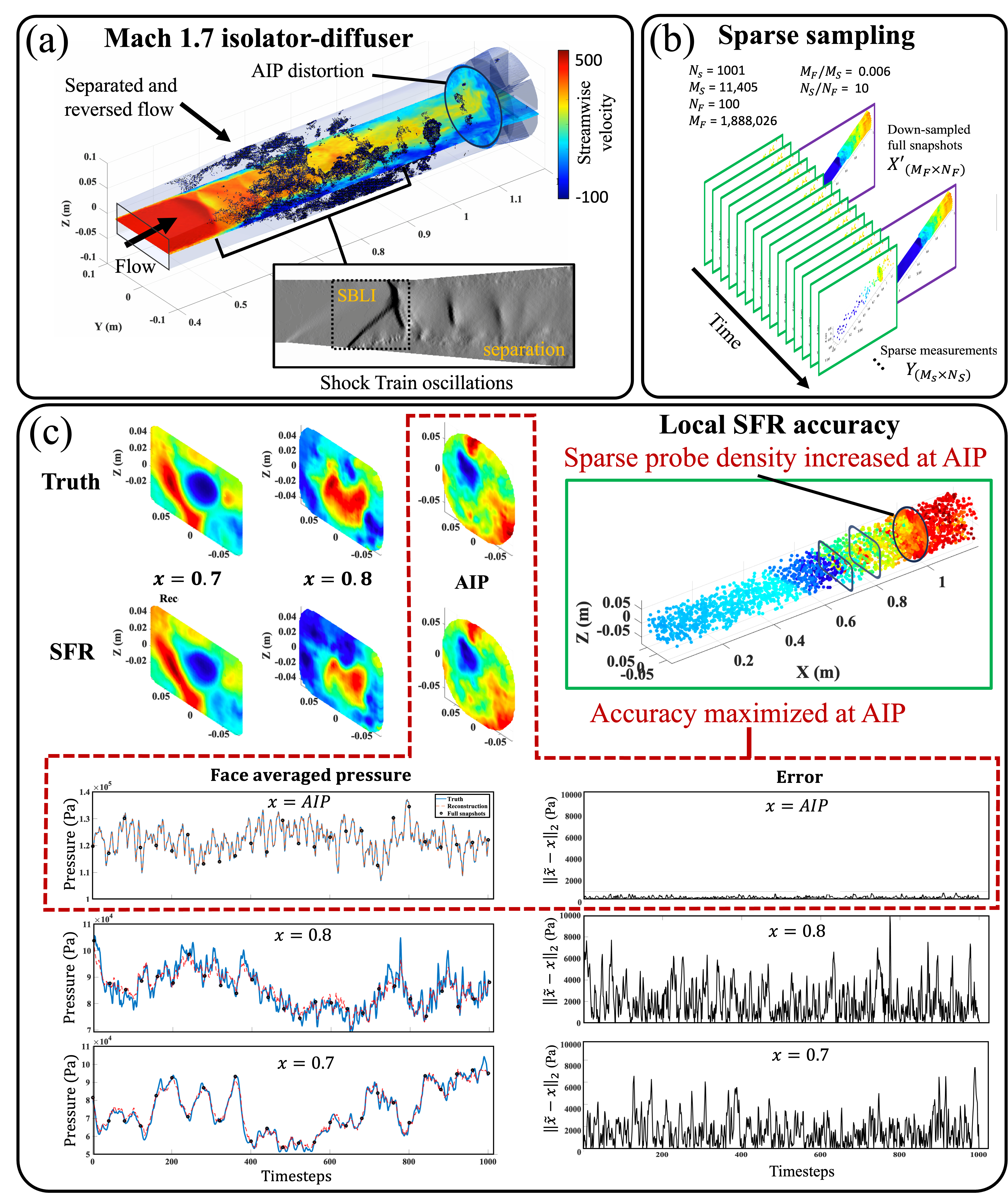}
\caption{(a) Instantaneous snapshot of the inlet isolator-diffuser flow path. Aerodynamic features such as the distortion at the AIP are identified. (b) Sparse sampling parameters for the local error example. (c) Instantaneous pressure distortion comparing the SFR and truth solutions at the $X=0.7, 0.8$ and the AIP cross sections. 
The face averaged pressure signals and error at these stations demonstrate how increase the sparse probe density at the AIP preserves accuracy with an order of magnitude comparison to the other planes.}
\label{fig:inlet_intro}
\end{figure}

The local SFR accuracy for the inlet distortion problem is discussed first to emphasize how judicious placement of sparse measurements can eliminate error in important locations, regardless of down-sampling parameters.
While adhoc, this is a valuable and practical feature for the preservation of rare distortion event dynamics at the AIP, exploited in Secn.~\ref{secn:CPOD}).
The current example returns to Fig.~\ref{fig:inlet_intro}(b) to reconstruct the previous $M_F=18m$ full snapshot solution, sampled at $f_S=40kHz$ for $N_S=1000$ snapshots; sufficient to demonstrate time-local behavior.
The chosen snapshot SFR compression rate is reduced to a moderate $C=10$, by changing the down-sampled full snapshot frequency to $f_F=4 kHz$; this reduction is based on an acceptable accuracy determined from the upcoming sensitivity study.




Local SFR accuracy is examined at certain axial stations in the the diffuser test section, where the SBLI induced separation distorts the flow approaching the AIP, as shown in Fig.~\ref{fig:inlet_intro}(c). 
The AIP cross-section (red block) is intentionally populated with an extra $9500$ sparse probes to increase accuracy, comprising about $5 \%$ of all sparse measurements throughout the duct in this example.
Note that in practice, perfect accuracy is recovered by overwriting the corresponding subset of the approximated solution, $\tilde{\textbf{X}}$, with the sparse measurements, $\textbf{Y}$; however, this was not done here to demonstrate how denser sparse measurements improve error in general. 
In Fig.~\ref{fig:inlet_intro}(c), the AIP pressure field is plotted, comparing instantaneous snapshots of the truth, $\textbf{X}$, and the SFR approximation, $\tilde{\textbf{X}}$, which is nearly identical.
The face-averaged pressure signals are also plotted with the down-sampled full snapshot instances circled for reference.
The local error norm, $||\tilde{x}_n-x_n||_2$, plotted as a function of time in the right column, confirms an almost exact reproduction of the full solution, effectively guaranteeing all extreme distortion events in the signal are captured.

For comparison, the instantaneous SFR accuracy of pressure at two cross-sections further upstream from the specified AIP probes is examined.
Figure~\ref{fig:inlet_intro}(c) shows these locations within the random distribution of probes at $x=0.7$ and $0.8$, both intersecting the heart of the unsteady shock train where fluctuations are most intense. 
The pressure fields show the truth and SFR approximation have good qualitative agreement in reproducing the turbulent spatial patterns, although the $x=0.8$ station is somewhat deficient, likely due to a lack of probes near this plane.
Error analysis of the spatially-averaged pressure signals at $x=0.7$ and $x=0.8$, are generally accurate, but have two orders of magnitude greater error than the previous AIP results.
Temporal accuracy is challenged most half-way between full snapshot intervals, but recovers and approaches zero error near coinciding full and sparse snapshot intervals.
This Nyquist-like sampling error is explored in greater detail later in Secn.~\ref{secn:proxy_Error}.

\subsection{Global accuracy study}

The global trade-off between accuracy and compression is now investigated as a function of SFR sampling parameters.
As will be shown, error is primarily driven by the disparity in temporal sampling rates between the full snapshots and sparse measurements, expressed as the interval sampling frequency ratio $f_F/f_S$, or used here, the ratio of full to sparse snapshots $N_F/N_S$.
That is, temporal sampling error decreases to zero as $N_F/N_S \rightarrow 1$. 
To a lesser degree, a smaller number of sparse measurements, $M_S$, also contributes to error, and is expressed relative to the size of the full snapshots, $M_S/M_F$.
By design, $M_S/M_F<<1$, but it is important to understand the minimum number of points to be included before accuracy starts to degrade for a CFD problem of this size.

The global error, $E$, associated with all sources of error is measured by the L2 norm difference between the SFR and truth snapshots, summed across all time-steps and normalized,
\begin{equation}\label{Eqn:global_error}
    E=\frac{\sum_{n}\| \tilde{\textbf{X}}_n - \textbf{X}_n \|_{2}}{\sum_{n}\| \textbf{X}_n\|_{2}},
\end{equation}
and is expressed as an accuracy percentage, $A$,
\begin{equation}
    A=100\times(1-E).
\end{equation}
To measure the temporal error behavior, the L2 norm difference at each time-step is also used, as done in Fig.~\ref{fig:inlet_intro}(c),
\begin{equation}
    E(t)=\| \tilde{x}_n(t) - x_n(t) \|_{2}.
\end{equation}

Since the error analysis requires the full "truth" snapshots ($\textbf{X}$) at a high-frequency, the sampling rate and duration for this example are reduced compared to the maximum SFR capability; this limitation is later relaxed with a technique to measure error using only the sparse measurements when truth snapshots are not available.
Here, the sparse sampling rate is fixed to $f_S=10kHz$ and $f_F$ is varied such that increasingly compressed sampling ratios of $N_F/N_S=f_F/f_S=1,0.5,0.2,0.1,0.02,0.01,$ and $0.0002$ are tested. 
The full streamwise velocity solution is reconstructed over $0.1$ seconds of data with $N_S=1000$ sparse samples.

\begin{figure}
\centering
\includegraphics[width=1\textwidth]{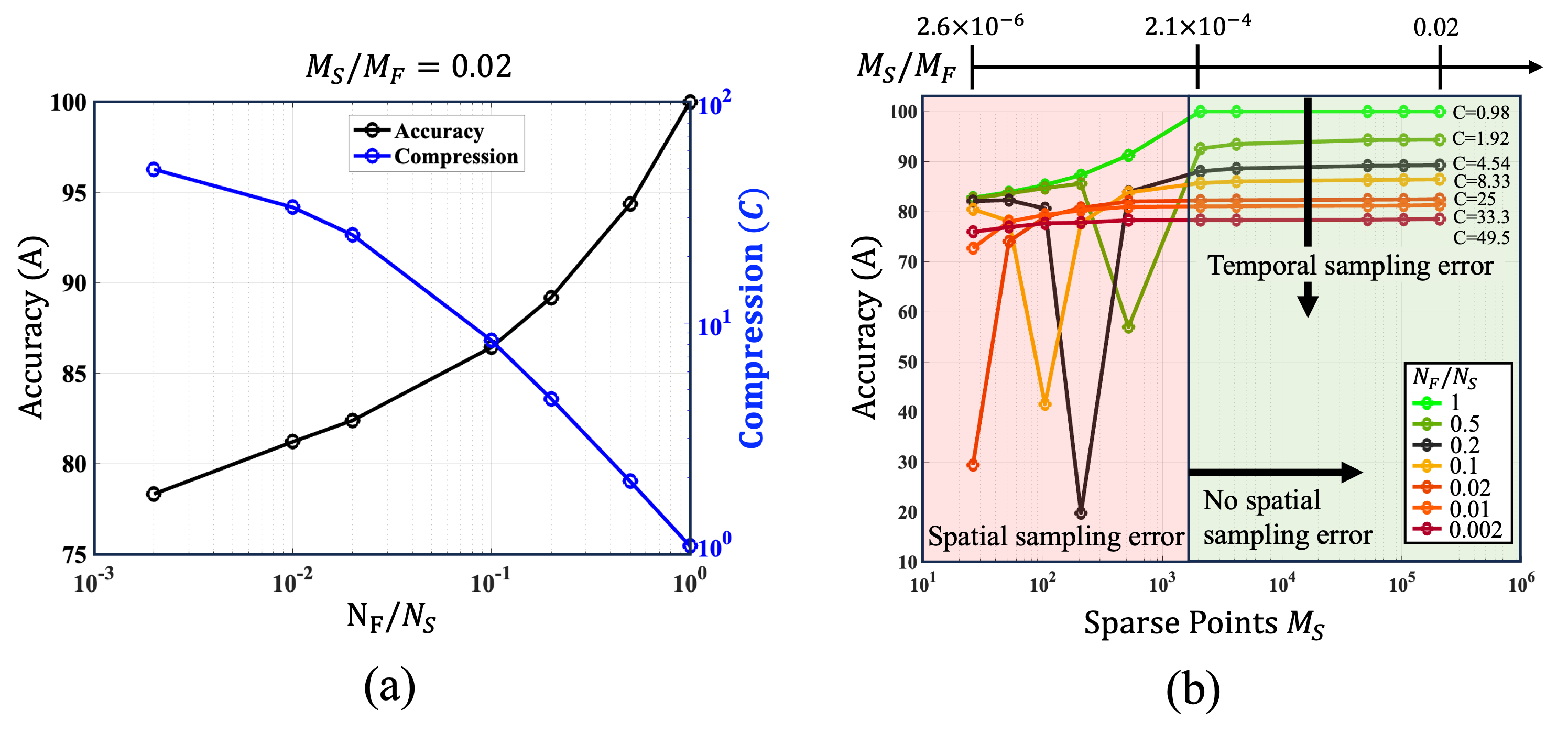}
\caption{(a) Compression and accuracy of the snapshot SFR method as a function of full snapshot sampling, $N_F/N_S$, for the largest number of sparse points $M_S/M_F=0.02$. (b) The accuracy as a function of both full snapshot sampling rates, $N_F/N_S$, and the number of sparse points relative to the full sized snapshots, $M_S/M_F$. 
Above $M_S/M_F > 2.1\times 10^{-4}$, the number of sparse probes does not contribute error, which is dominated by the temporal sampling rates.}
\label{fig:tot_error}
\end{figure}

Global accuracy as a function of temporal sampling, $N_F/N_S$, is presented in Fig.~\ref{fig:tot_error}(a) using all available sparse samples ($M_s=207k$, $M_S/M_F=0.02$), followed by a study in (b), which also decreases the sparse points from the maximum $M_S/M_F=0.02$ available, to a minimum of $2.6\times 10^{-6}$ for each value of $N_F/N_S$.
Figure~\ref{fig:tot_error}(a) displays the basic inverse relationship between compression and accuracy.
Accuracy improves monotonically when $N_F/N_S$ increases, reaching perfect accuracy when $N_F/N_S=1$.
An important result is at $N_F/N_S=0.5$, which is the minimum sampling interval that collects every other full snapshot to double the temporal resolution of the LES.
This nominal $2\times$ compression only marginally affects error, with a global accuracy drop to $A=94\%$.
In the extreme compression limit of $N_F=3$ full snapshots ($N_F/N_S=0.002$, $C=49.5$), a fair accuracy of $77\%$ is obtained.
Information associated with detailed fluctuations will of course be significantly impacted, since the unsteady solution is effectively being interpolated across a combination of only 3 snapshots.
However, this demonstrates a clear benefit of the SFR method, that the accuracy is well bounded regardless of the sampling rate, a property of having even a small basis represent a subset of the truth solution. 

The secondary source of error comes from the number of sparse measurement locations, $M_S$, and is examined in Fig.~\ref{fig:tot_error}(b) for each temporal sampling ratio, $N_F/N_S$.
At the maximum number of sparse points used ($M_S/M_F=0.02$), the accuracy reproduces the results of (a), and are well into the stable regime (green box), where accuracy is constant up to the threshold of decreasing the sparse measurements below $M_S/M_F=2.1\times 10^{-4}$.
Beyond this limit, accuracy deteriorates, sometimes gradually, and occasionally with massive error.
The key finding from Fig.~\ref{fig:tot_error}(b), is that above this unstable regime, defined at approximately $\textbf{$M_S/M_F \approx 1 \times 10^{-3}$}$ (or $0.1\%$ the size of the full snapshot), the number of sparse points effectively contributes no additional error, demonstrating that SFR will preserve accuracy to the maximum that the temporal sampling will allow, determined foremost by $N_F/N_S$.
Of course, the exact threshold will change with each problem, and the number of sparse points could be further reduced with sensor optimization.
However, the cheap writing cost of an even more conservative $1\%$ dedication to sparse measurements, justifies the random brute force approach, particularly when combined with judicious placement in important local regions of the flow.





\begin{figure}
\centering
\includegraphics[width=1\textwidth]{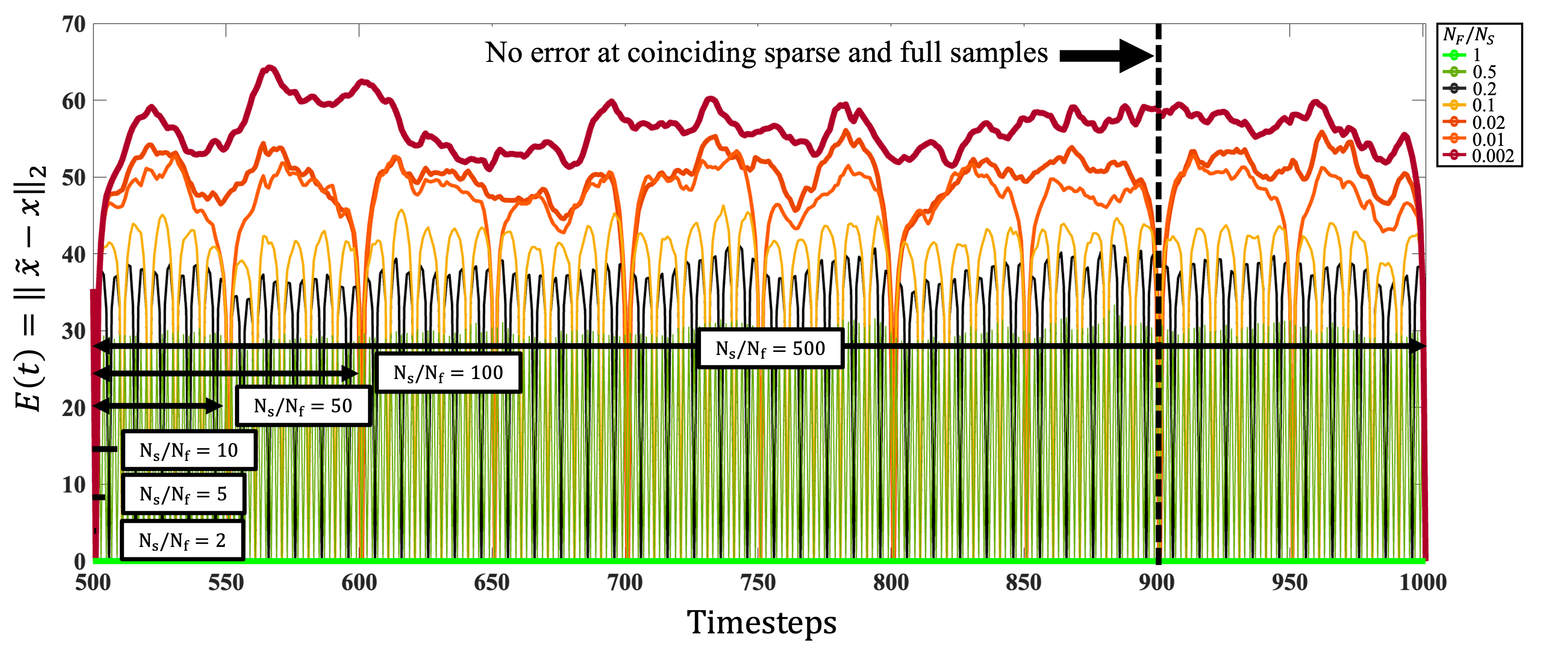}
\caption{The transient error norm of the snapshot SFR method for different full snapshot sampling rates $N_F/N_S$. Error is zero at intervals when full and sparse snapshots coincide, and reaches a maximum between these instances, increasing as the full snapshots become fewer.  }
\label{fig:transient_error}
\end{figure}

Next in Fig.~\ref{fig:transient_error}, the global error is investigated as a function of time, $E(t)$.
The maximum number of sparse points is used ($M_S/M_F=0.02$) for all cases to eliminate this source of error, as concluded from the earlier results of Fig.~\ref{fig:tot_error}(b).
Instead, the temporal error better relates to the parameter, $N_F/N_S$, which is again tested using the interval sampling approach ($f_F/f_S=N_F/N_S$). 
Over the $N_S=1000$ timesteps reconstructed, the transient error varies from zero at intervals when both sparse and full snapshots coincide, i.e. every $N_S/N_F$ timesteps, to a maximum error half-way between full snapshots.
The longer the duration between full snapshots, the greater the peak transient error becomes. 
If this duration exceeds any correlated time scale in the flow, i.e. full snapshots that are independent from one another, the error saturates to a certain level, as seen with cases $N_S/N_F > 100$.
On the other hand, a higher-frequency full snapshot sampling rate, $f_F$, can bias error at that time-scale, aliasing the SFR approximation to a degree dependant on the disparity between sampling rates $f_F/f_S$.
Therefore, this type of Nyquist-like frequency error should be noted when using the SFR method.
Next, it is shown how this temporal error manifests as a loss of energy in the solution between full snapshot intervals, and how this aliasing can be diminished using a consecutive sampling approach.

\subsection{Sparse error proxy and sampling methods}\label{secn:proxy_Error}




As a feature of the SFR framework, we demonstrate a method to estimate global and temporal accuracy using sparse measurements as a proxy for error throughout the full domain, providing a performance metric that does not require the self-defeating purpose of writing the full truth solution. 
This feature works because the input sparse snapshots, $\textbf{Y}$, are a subset of the truth solution, \textbf{X}, and the sparse subset of the reconstructed solution, $\tilde{\textbf{Y}}$, is subjected to reconstruction error.
Thus an accuracy metric from this difference can be made that follows the full domain error trends.
The sparse measurement error, $E_Y$, is
\begin{equation}
    E_Y=\frac{\sum_{i}\| \tilde{\textbf{Y}}_i - \textbf{Y}_i \|_{2}}{\sum_{i}\| \textbf{Y}_i\|_{2}},
\end{equation}
or as an accuracy percentage, $A_Y$,
\begin{equation}
    A_Y=100\times(1-E_Y).
\end{equation}
This sparse error calculation is also a much smaller and faster calculation than the global error of Eqn.~\ref{Eqn:global_error}.
Note that after this error calculation, the subset of sparse measurements in the reconstructed solution, $\tilde{\textbf{Y}}$, should ultimately be substituted with the input sparse measurements, $\textbf{Y}$, since they are freely available and have no error.



\begin{figure}
\centering
\includegraphics[width=1\textwidth]{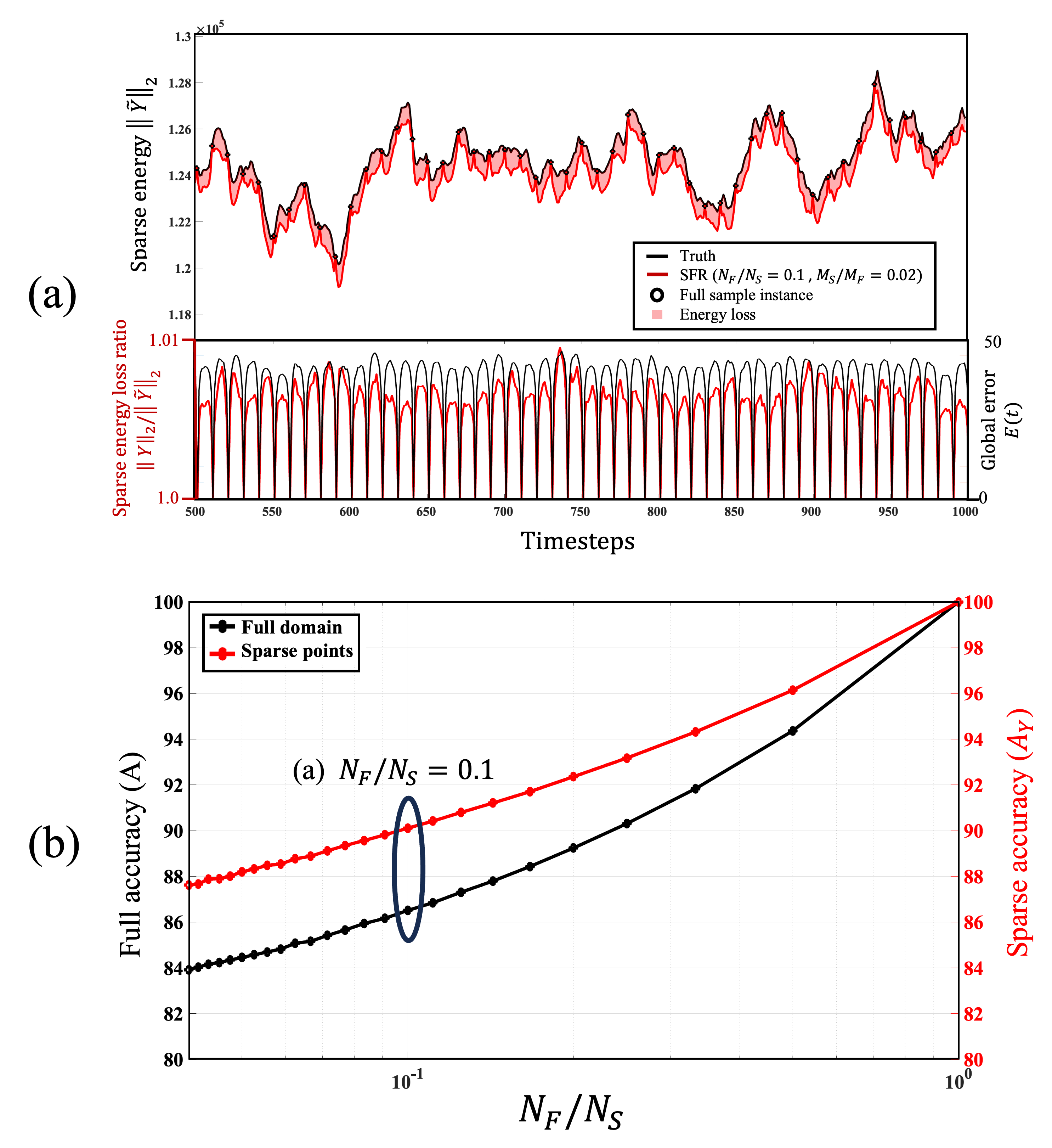}
\caption{(a) Sparse measurement energy as a function of time comparing the truth and SFR ($N_F/N_S=0.1$ and $M_S/M_F=0.02$) results, showing the error manifests as an energy loss and follows the global frequency-biased error of the full snapshots. (b) Comparison of sparse measurement and full snapshot accuracy metrics, $A$, and $A_Y$, respectively, as a function of $N_F/N_S$, validating the use of sparse measurements as a proxy for global error. }
\label{fig:sparse_transient}
\end{figure}

Results using the sparse measurement proxy are reported in Fig.~\ref{fig:sparse_transient} to demonstrate how temporal error manifest as energy loss in the form of weaker fluctuations, as measured by the L2 norm. 
This example uses moderate SFR parameters of $N_F/N_S=0.1$ and $M_S/M_F=0.02$, with a compression ratio of $C=8.33$.
Fig.~\ref{fig:sparse_transient}(a) shows the sparse energy for the truth solution and SFR approximation, $||\tilde{\textbf{Y}}||_2$, with their difference, $||\tilde{\textbf{Y}}-\textbf{Y}||_2$, representing the error and energy loss. 
Also shown in Fig.~\ref{fig:sparse_transient}(a), is the sparse energy loss ratio, $||\tilde{\textbf{Y}}||_2 /||\textbf{Y}||_2$, compared to the global temporal error, $E(t)$, demonstrating the sparse error (and energy loss) follows the frequency of the full snapshot sampling error, as with Fig.~\ref{fig:transient_error}.
The global accuracy over all time is plotted in Fig.~\ref{fig:sparse_transient}(b), comparing the full, $A$, and sparse, $A_Y$, accuracy as a function of $N_F/N_S$.
The $N_F/N_S=0.1$ case in (a) is circled and highlights a $4 \%$ discrepancy in the sparse accuracy and the entire solution.
Overall, the sparse error proxy over-predicts, but closely parallels, the trends of the full domain accuracy, validating its use to estimate global and temporal error in the reconstruction.
A similar relation between $A$ and $A_Y$ follows as a function of $M_S/M_F$, but is not reported since this error is primarily contributed by $N_F/N_S$, as demonstrated previously in Fig.~\ref{fig:tot_error}.

\begin{figure}
\centering
\includegraphics[width=1\textwidth]{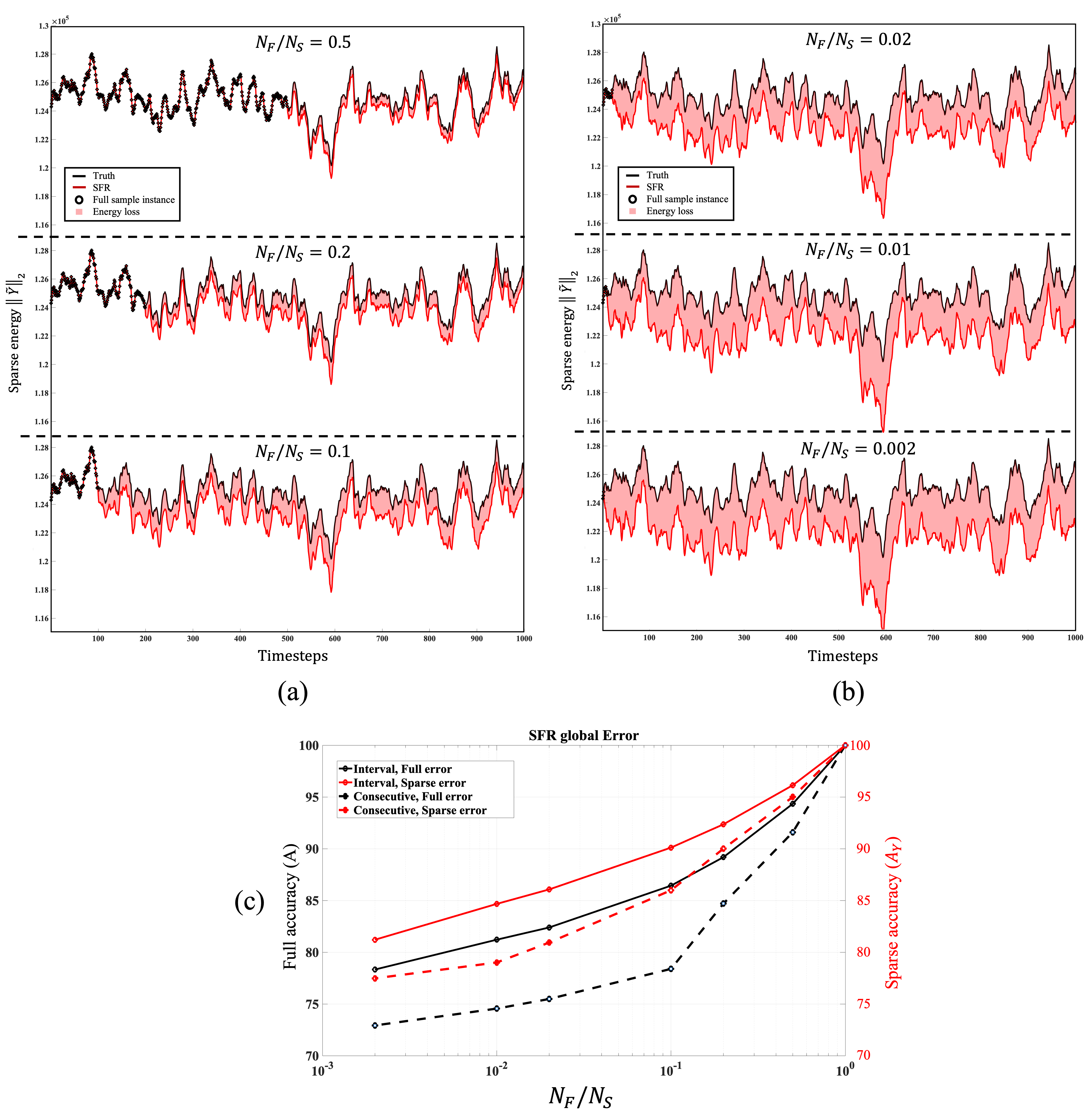}
\caption{(a-b) Temporal sparse energy loss and error for the consecutive, up-front, full snapshot sampling approach for a decreasing ratio of full snapshots $N_F/N_S$. Accuracy is still preserved at coinciding full and sparse snapshot instances. Error grows in the latter extrapolation portion where only sparse measurements are taken, but is free from the frequency aliasing energy loss behavior of the interval sampling method. (c) Global error accuracy measured at sparse and full domain shows the overall error is larger for the consecutive sampling method for all $N_F/N_S$ cases. }
\label{fig:consec_vs_interval}
\end{figure}

Given the effectiveness of the sparse measurement error to reflect energy loss as a function of time, these metrics are used again to compare different SFR full snapshot sampling approaches.
Specifically, the frequency-based "interval" approach with down-sampled full snapshots coinciding with sparse measurements (as used so far), is compared to the "consecutive" method that collects full snapshots up-front at the higher frequency $f_S$.
This latter approach could be considered an initial learning phase of the CFD solution, where the desired $N_F$ snapshots are written up-front, followed by only writing $N_S$ sparse measurements for the remainder of the simulation.
Figure~\ref{fig:consec_vs_interval} tests this sampling approach for decreasingly smaller ratios of $N_F/N_S$, in (a-b) as a function of time, and in (c) globally.
The same data and number of sparse measurements is used here as in Fig.~\ref{fig:transient_error}.
Across all cases in (a) and (b), a fundamental principle holds, that accuracy is perfectly preserved at intervals when full snapshots coincide with sparse snapshots.
Front-loaded to the first half of the solution in $N_F/N_S=0.5$, we observe the remainder of the time is subject to a degree of error and energy loss.
A notable benefit is that unlike the interval frequency sampling approach in Fig.~\ref{fig:sparse_transient}(a), the consecutive snapshot error here is free from aliasing, where energy losses oscillate at the full snapshot sampling frequency $f_F$. 
When front-loaded, the error is evenly distributed throughout the remaining time, but grows significantly as $N_F/N_S \rightarrow 0$, as shown for all cases in (b).
The global accuracy in Fig.~\ref{fig:consec_vs_interval}(c) indicates that the overall error is however worse for the consecutive method than the interval method, consistent for both the full domain and sparse proxy accuracy metrics, $A$ and $A_Y$, respectively.
This is attributed to each SFR sampling method being interpreted as an interpolation or extrapolation problem, of which the latter is more difficult. 




\section{POD-SFR Performance}\label{secn:POD_perf}




The performance of the POD-SFR method is now tested to evaluate the accuracy and increased compression of using fewer POD modes as the basis, $\textbf{B}$, instead of full-sized snapshots.
For this method, POD modes can be directly computed from the available snapshots $\mathbf{X}'$, as discussed in Secn.~\ref{secn:POD_SFR_method}.
Alternatively, the "double POD-SFR" method (Secn.~\ref{secn:mem_improvements}) is showcased here to circumvent the memory limitations ($128 GB$ of RAM) that prevented the calculation of all full sized POD modes.
As with the snapshot SFR results, the streaming variant is also used in conjunction with the interval samping approach.

\begin{figure}
\centering
\includegraphics[width=1\textwidth]{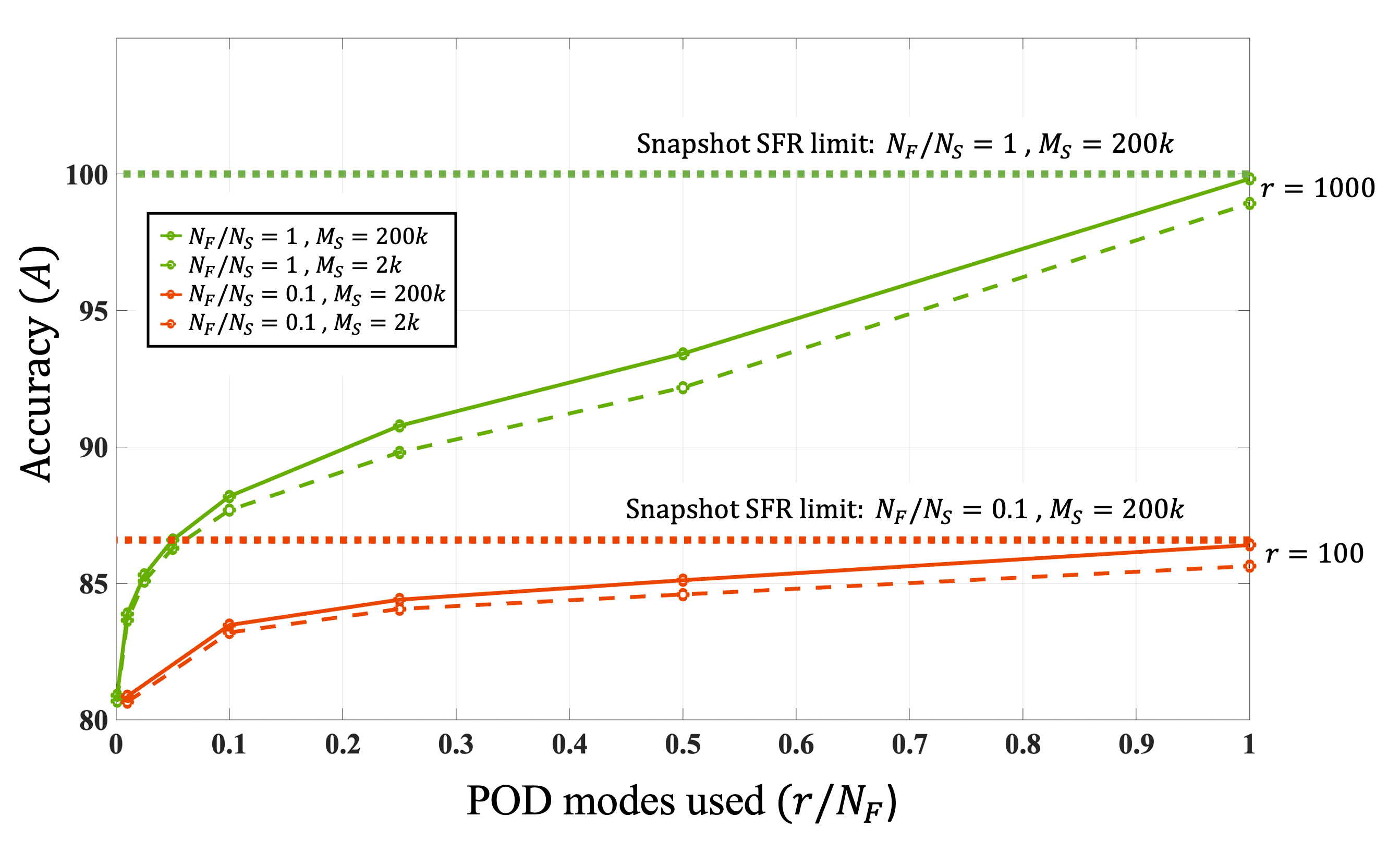}
\caption{POD-SFR accuracy of the $N_F/N_S=1$ baseline and $N_F/N_S=0.1$ cases as a function of the number of truncated POD modes used $r/N_F$. The dashed horizontal lines are the maximum accuracy possible using the snapshot SFR method, approached when all POD modes are used. When two orders of magnitude less sparse points are used ($M_S=2k$), the accuracy is only marginally effected. }
\label{fig:E_POD_sweep}
\end{figure}

Global accuracy results of four cases are reported in Fig.~\ref{fig:E_POD_sweep} as a function of the number of POD modes kept relative to those available ($r/N_F$).
The degree of sparsity chosen, $N_S/N_F$ and $M_S/M_F$, is guided by the results in Fig.~\ref{fig:tot_error}(b), selecting cases to best demonstrate the properties of the POD-SFR method.
First examined is the $N_F/N_S=1$ case, where full and sparse snapshot sampling rates are equivalent and there is no data compression.
This serves as a baseline to ensure the POD-SFR solution recovers the snapshot SFR and truth solution in this limit. 
For the second case, $N_F/N_S=0.1$ is chosen again for its moderate compression ($C=8.33$) and accuracy trade-off ($A=86.5\%$).
These two cases use the largest number of sparse points tested ($M_S=200k, M_S/M_F=0.02$), where spatial sampling error was found negligible, and are compared with two more cases, where the number of sparse probes is lowered by two orders of magnitude ($M_S=2k, M_S/M_F=2.1\times 10^{-4}$), just above the threshold where meaningful error is introduced.

The horizontal dashed lines in Fig.~\ref{fig:E_POD_sweep} mark the maximum possible accuracy obtained from the snapshot SFR approach for the given sampling ratio. 
All cases converge to this level of accuracy as more POD modes are added, almost exactly recovering the snapshot SFR results when $r/N_F=1$ in both cases.
Both sampling rates, $N_S/N_F=1$ and $0.1$, start with the same accuracy at $81\%$ when a single $r=1$ POD mode is used, since both are using the mean-flow as the basis. 
However, the accuracy rises much quicker for $N_S/N_F=1$, because there are more POD modes to expand the basis derived from more full snapshots; at $r/N_F=1$, there are $1000$ POD modes, compared to the $100$ POD modes of the $N_F/N_S=0.1$ case.
Regarding a decrease in the number of sparse measurement locations, $M_S$, the accuracy is only mildly affected for both cases.

\begin{figure}
\centering
\includegraphics[width=1\textwidth]{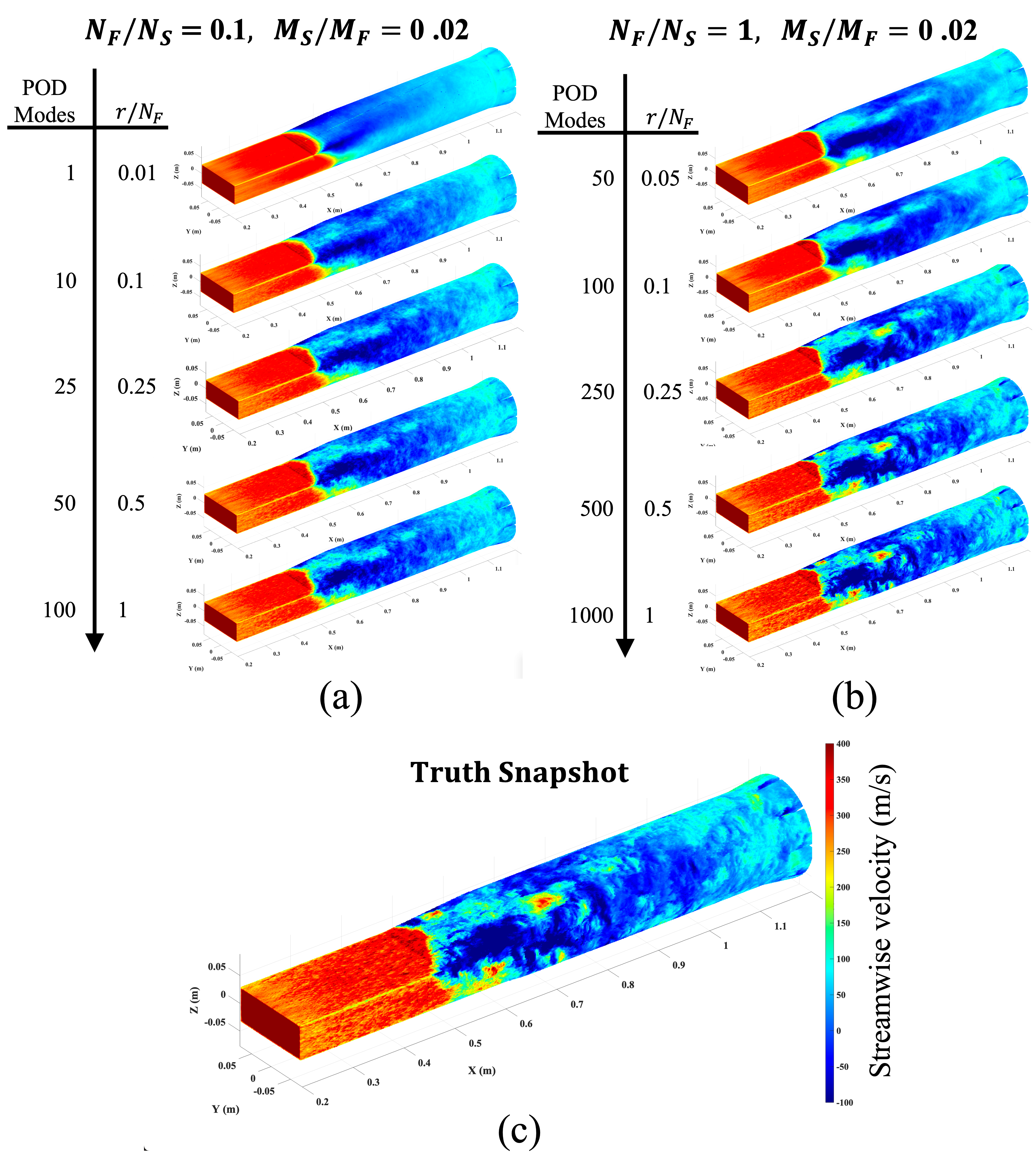}
\caption{Reconstruction of an instantaneous full snapshot for the (a) $N_F/N_S=0.1$ and (b) $N_F/N_S=1$ cases, compared to the (c) truth snapshot. The qualitative features improve as more POD modes, $r/N_F$, are added to the POD-SFR basis.}
\label{fig:POD_instant}
\end{figure}

Reconstructed instantaneous velocity snapshots are compared in Fig.~\ref{fig:POD_instant} for a qualitative illustration of POD-SFR compression effects on the flow.
The number of POD modes kept in the basis relative to the total number of available full snapshots ($r/N_F$), is compared for the (a) $N_F/N_S=0.1$ case, (b) $N_F/N_S=1$ case, which provides the maximum accuracy the POD-SFR can produce, and (c) the truth snapshot at that instant.
Even with a few modes, the dominant flow structures emerge in both cases, but with diminished turbulent fluctuation amplitudes, recovering as more modes are added.
Of particular note is the nearly indistinguishable results between $r/N_F=0.5$ and $r/N_F=1$ for both cases, indicating that an additional $50\%$ compression can be obtained for a given down-sampling with little qualitative differences in the results.
Figure~\ref{fig:POD_instant} also demonstrates that for the same number of POD modes derived from a small or larger set of snapshots (for example, $r=50$ from $r/N_F=0.5$ and $0.05$), the results can be just as effective in producing the same qualitative observations about the flow, but with significant additional compression rates.
Finally, consider implementation of the double POD-SFR method combined with the streaming variant.
Changing the number of POD modes used in the basis, $r$, provides an efficient way to throttle accuracy when analyzing large datasets, exchanging memory cost for the speed and resolution of the current snapshot calculation in real time.





\section{Preservation of extreme event dynamics}\label{secn:CPOD}

\subsection{CST-POD inlet distortion events}

Studying transient flow events can be data intensive, especially for rarely occurring events which require a prolonged time series at high sampling rates to statistically converge.
As previously demonstrated, SFR enables a more efficient workflow for analyzing such data. 
However, appropriate down-sampling of parameters must be tested to ensure the approximation preserves not only global accuracy, but local event dynamics, particularly when the event timescales are faster than the full snapshot sampling rate.
The conditional space-time decomposition (CST-POD) has proven effective at characterizing rare events \cite{Schmidt2019,stahl_2023_CPOD_jcp} and is used here to evaluate event reconstruction accuracy.
Resulting CST-POD modes are ranked by energy and produce an ensemble average of all events for the first mode, with non-leading modes delineating space-time correlation across different event types or underlying dynamics.
In this section, a CST-POD analysis is conducted, comparing the SFR and truth datasets to determine if all event dynamics are faithfully reproduced as a function of temporal, spatial, and modal sampling parameters, $N_F/N_S$, $M_S/M_F$, and $r/N_F$, respectively.

\begin{figure}
\centering
\includegraphics[width=1\textwidth]{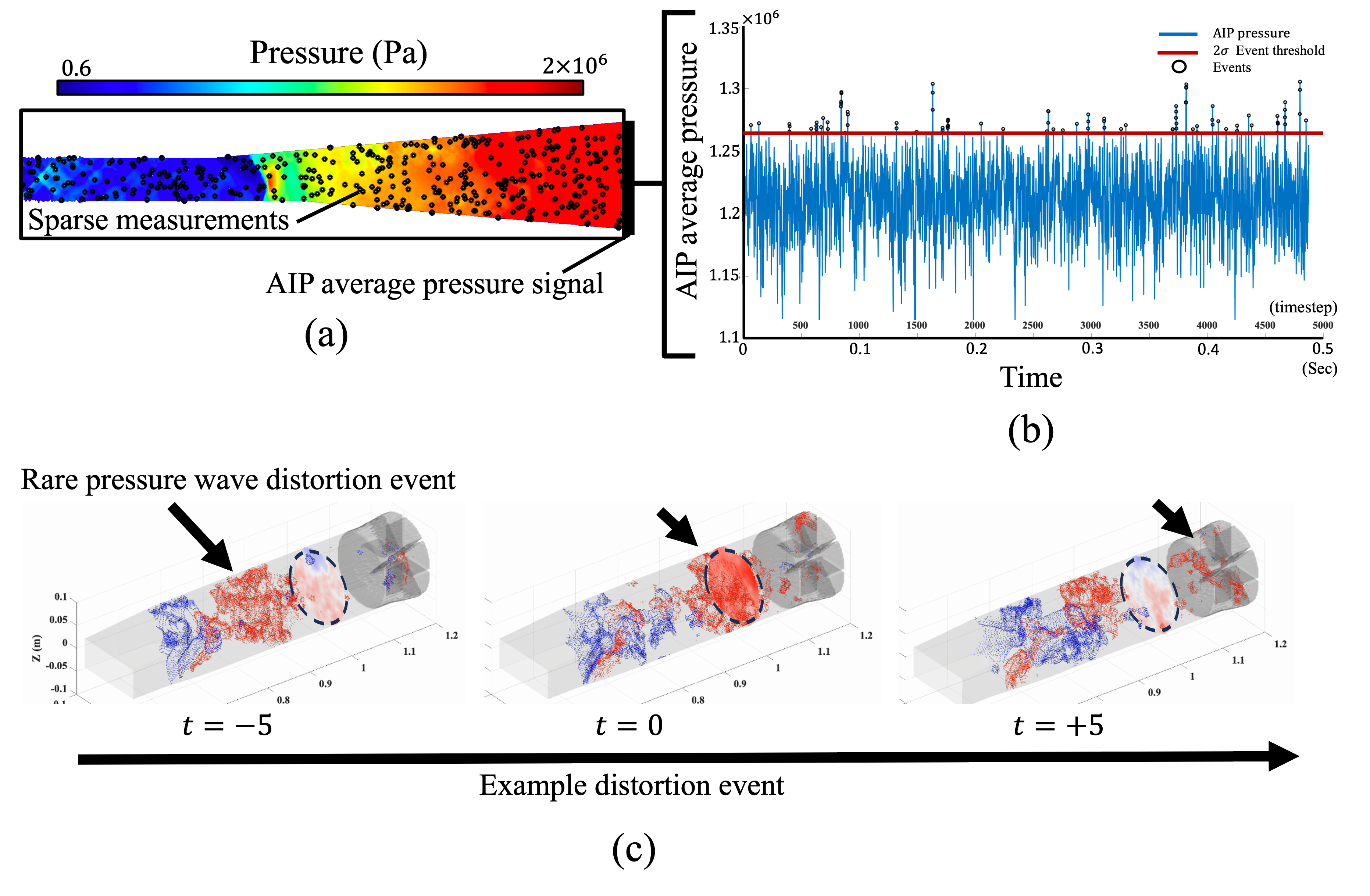}
\caption{CST-POD analysis of extreme and rare distortion events. (a) 2D pressure field of the inlet showing sparse measurements locations and the AIP where distortion events are measured. (b) Pressure signal averaged across the AIP, with extreme fluctuation events identified by a 2 standard deviation threshold. (c) An example of a high-pressure distortion event passing through the AIP. }
\label{fig:single_event}
\end{figure}

For the transonic inlet example, the rare events are the extreme pressure spikes measured at the exit of the diffuser (the aerodynamic interface plane, AIP).
These distortion events are instantiated in Fig.~\ref{fig:single_event}, showing the (a) center-plane pressure field with sparse measurement locations and the (b) AIP location where unsteady pressure is spatially averaged across.
Recall, the sparse probe density is increased at the AIP, guaranteeing the signal has maximum accuracy and that all events are identified, regardless of SFR parameters.
The CST-POD method identifies the high pressure events from this signal in Fig.~\ref{fig:single_event}(b), determined by pressure fluctuations greater than two standard deviations of the signal.
Since the pressure signal is averaged over the AIP, many unique events are captured across all realizations, the correlated trends of which can be separated by CST-POD modes.
A representative distortion event is shown in Fig.~\ref{fig:single_event}(c), where a large red coherent structure, first observed at $t=-5$ snapshots before the event, crosses the AIP at the moment of the pressure spike ($t=0$), then convects downstream and interacts with the instrumentation rake at $t=+5$.
This entire event lasts approximately $0.001$ seconds.
For the CST-POD analysis, a total of 89 events were identified over a $0.5$ second period, collecting two-dimensional snapshots over an event window extending from $t=-40$ to $t=+10$ snapshots so that important causal dynamics preceding the event are captured.
Both the SFR and CST-POD calculations are performed on the two-dimensional pressure field for this example.

\begin{figure}
\centering
\includegraphics[width=1\textwidth]{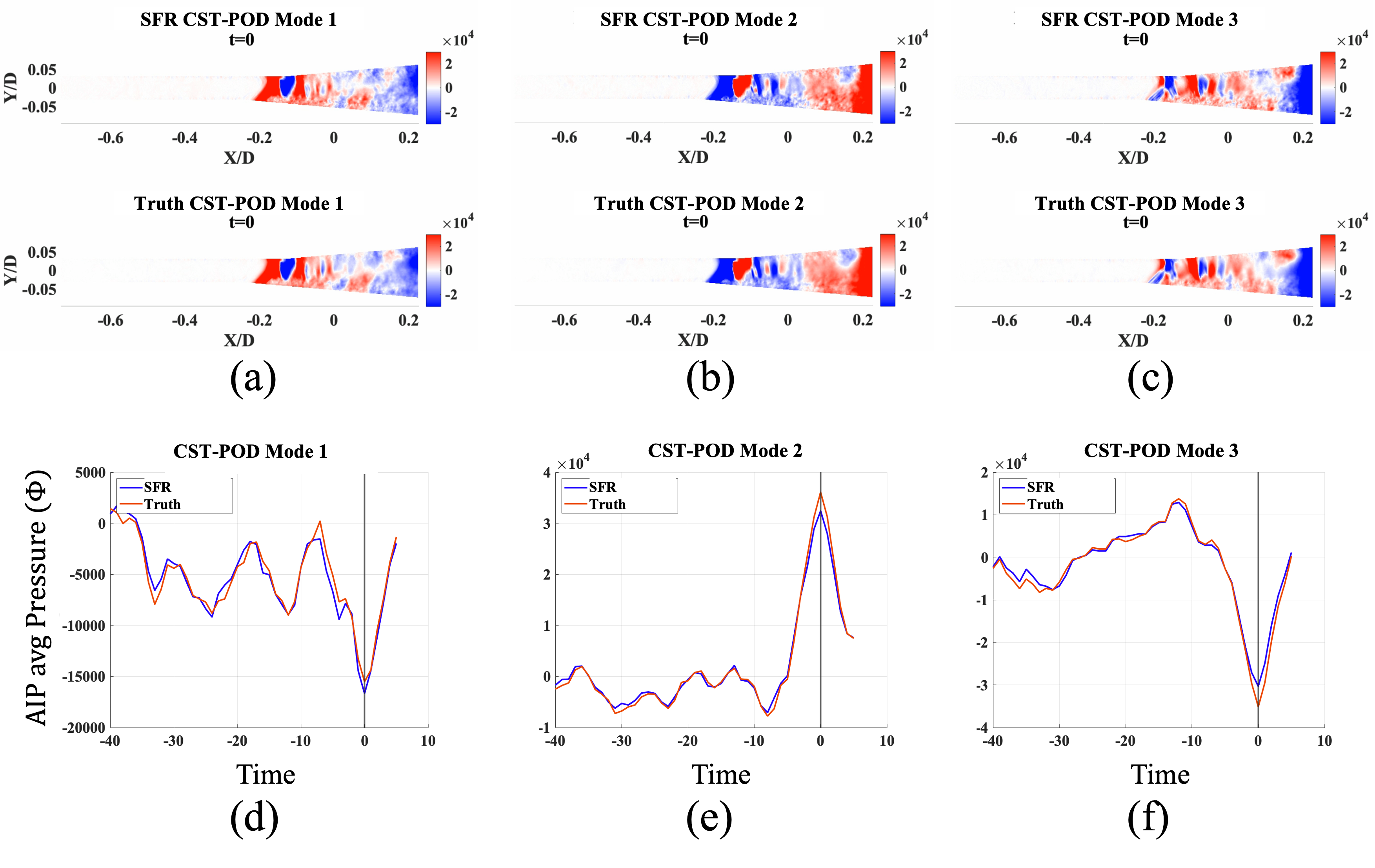}
\caption{Comparison of the CST-POD modes with the truth data and SFR data sampled at $N_F/N_S=0.05$ and $M_S/M_F=0.063$. (a-c) CST-POD mode instances at the moment of the distortion pressure spike event ($t=0$), and (d-e) corresponding pressure at the AIP throughout the CST-POD event duration.  }
\label{fig:CPOD_modes}
\end{figure}

Leading CST-POD modes from the SFR and truth data are compared in Fig.~\ref{fig:CPOD_modes}.
For demonstration purposes, optimally determined snapshot SFR parameters of $N_F/N_S=0.05$ and $M_S/M_F=0.063$ are initially used, yielding a compression ratio of $C=17.75$; a sensitivity study is conducted later.
The first three CST-POD modes are presented in Fig.~\ref{fig:CPOD_modes}(a-c), comparing the SFR and truth input data at the moment of the pressure spike event ($t=0$).
The corresponding transient behavior of each mode is also compared in Fig.~\ref{fig:CPOD_modes}(d-e), plotting the pressure correlation signal measured at the AIP.
Overall, the SFR extreme distortion events are nearly identical in both their spatial and temporal behavior for all three modes.
This suggest, at least for a lower rank two-dimensional dataset, a considerable SFR compression can accurately capture rare event dynamics.

\begin{figure}
\centering
\includegraphics[width=1\textwidth]{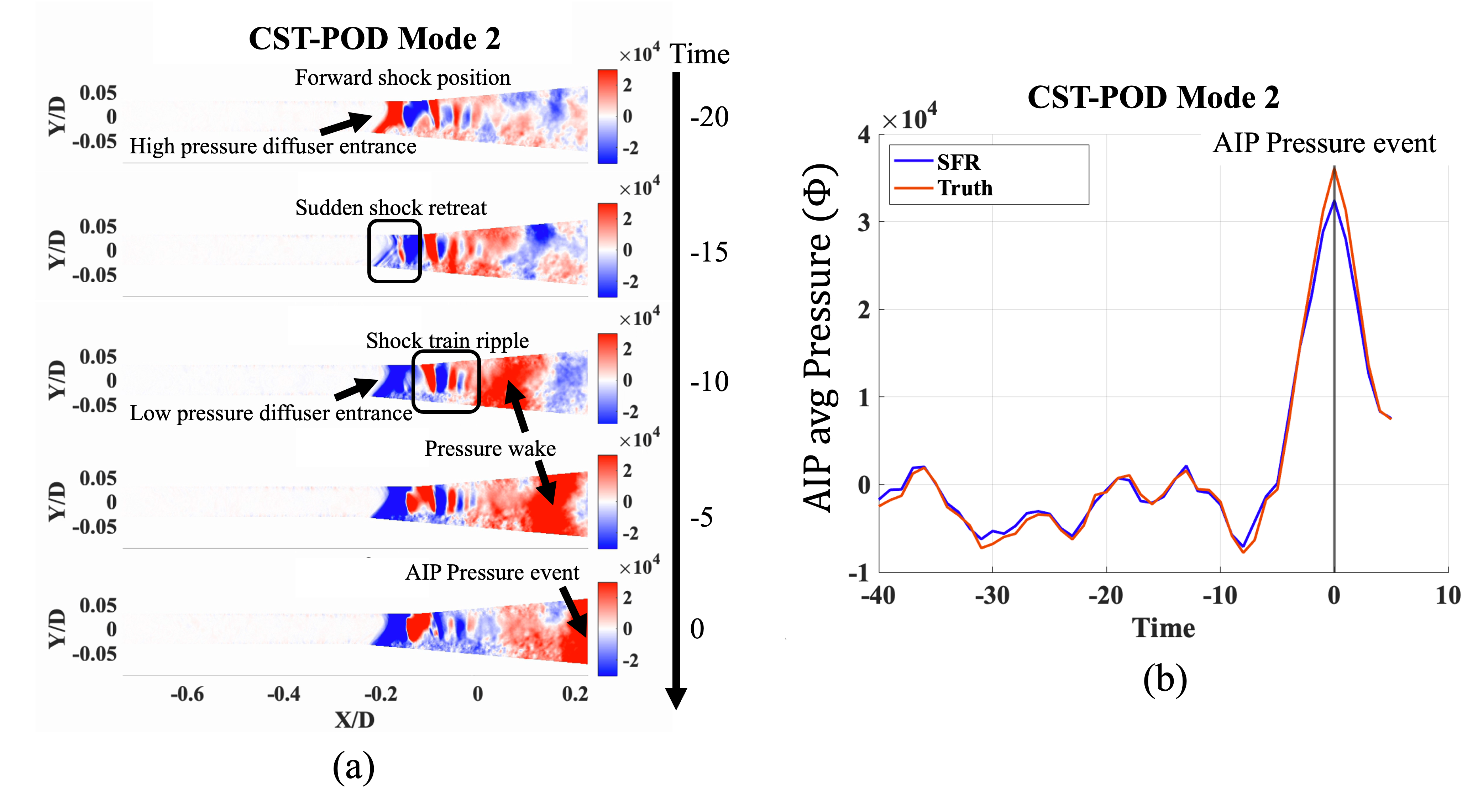}
\caption{CST-POD mode 2 (a) spatial evolution and (b) AIP pressure signal throughout the distortion event. A sudden retreat of the forward shock causes a pressure wave to ripple through the shock train and convect through the AIP a short time later. }
\label{fig:CPOD_mode2}
\end{figure}

For completeness, the entirety of CST-POD mode 2 is examined in Fig.~\ref{fig:CPOD_mode2}(a), along with the (b) corresponding AIP pressure signal to gain physical insight into the causation of the pressure distortion event.
Mode $2$ is selected over the CST-POD mode 1 ensemble average, which smooths over dynamic information when there is a wide variation in events \cite{stahl_2023_CPOD_jcp}.
Initially (before $t<-20$) the leading shock is in its forward position and the downstream pressure fluctuations near the AIP are relatively calm. 
At $t=-15$, a sudden retreat of the forward shock causes a large pressure wave to ripple through the shock train. 
This high pressure wake convects downstream passing through the AIP during the moment of the pressure spike event at $t=0$.

\subsection{Sensitivity study}

Regarding the SFR compression performance, a more quantitative assessment of event accuracy is presented by calculating the correlation (dot product) between all SFR and truth CST-POD modes.
\begin{figure}
\centering
\includegraphics[width=1\textwidth]{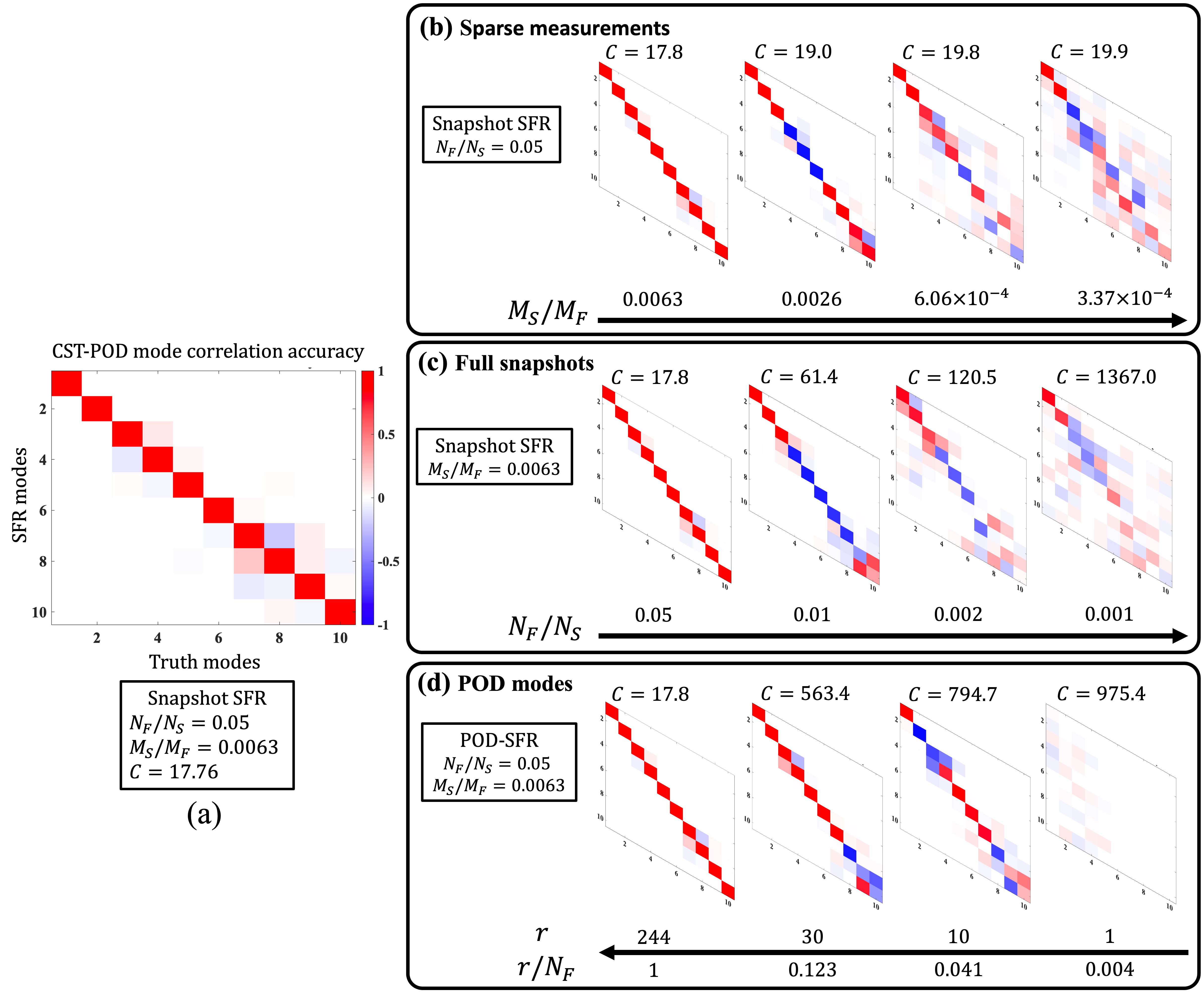}
\caption{SFR parameter sensitivity study that measures the accuracy of event dynamics by correlating the SFR and truth CST-POD modes. (a) CST-POD mode correlation matrix for the optimal case used in Fig.~\ref{fig:CPOD_modes}. Correlation matrices with further SFR compression of the (b) ratio of sparse probes ($M_S/M_F$), the (c) ratio of full snapshots ($N_F/N_S$), and the (d) ratio of POD modes ($r/N_F$) for the POD-SFR method.  }
\label{fig:CPOD_matrix}
\end{figure}
Figure~\ref{fig:CPOD_matrix}(a) exemplifies the previous SFR parameter results ($N_F/N_S=0.05$ and $M_S/M_F=0.063$) as a correlation matrix calculated across all SFR and truth modes. 
Attention is drawn to the mode pairs along the diagonal, with emphasis placed on the first few modes which carry the most energy and are therefore typically most important. 
Off-diagonal correlation pairs should ideally have a value of zero (white) since CST-POD modes are orthogonal.
A correlation of 1 (red), signifies the SFR modes are an exact match with the truth modes, while a correlation of -1 (blue) is interpreted as a flipped match (positive fluctuations are negative); this is a benign symmetry property of the singular value decomposition, but still indicates some SFR deviation.
Meaningful error is incurred when the correlation approaches zero on diagonal indices, while any correlation in off-diagonal modes suggests SFR accuracy is deteriorating with time-local event error spreading across modes.
For the best performing snapshot SFR parameters in Fig.~\ref{fig:CPOD_matrix}(a), all leading ten CST-POD modes display excellent agreement, with a near perfect correlation to the truth modes as demonstrated earlier in Fig.~\ref{fig:CPOD_modes}.

An accuracy sensitivity study is presented in Fig.~\ref{fig:CPOD_matrix} by further compressing the (b) ratio of sparse probes ($M_S/M_F$), the (c) ratio of full snapshots ($N_F/N_S$), and the (d) ratio of POD modes ($r/N_F$) for the  POD-SFR method. 
The degree of sparsity for each parameter is increased until meaningful changes are observed, while keeping all other parameters used in (a) constant.
In general for all parameters, anti-correlated mode-pairs begin to appear after another order of magnitude further compression, with off-diagonal information loss starting at the higher, less energetic, CST-POD modes. 
With another order of magnitude compression, only the first few modes remain accurate, confirming that at least the ensemble average is conserved (mode 1).
In the highest compression ratio limits, even these modes struggle to maintain any correlation to the truth and accuracy breaks down.

Detailed results from the CST-POD sensitivity study in Fig.~\ref{fig:CPOD_matrix} for each SFR compression parameter are summarized as follows:
\begin{enumerate}
    \item \textbf{SFR number of sparse measurements ($M_S/M_F$)}:  An approximately $1\%$ allocation of data to sparse measurement locations sufficiently maximizes accuracy. The $M_S/M_F=0.63 \%$ case gives an optimal compression of $C=17.76$ while still preserving rare event dynamics. At half that number of sparse points, nonleading modes degrade. Beyond this marginal accuracy, all results are severely effected. 
    \item \textbf{SFR number of full snapshots ($N_F/N_S$)}: Reducing the ratio of full snapshots to sparse snapshots produces reliable results up to $N_F/N_S=0.05$. However, noting the larger three-dimensional results from Secn.~\ref{secn:snapshot_results}, the temporal sampling error is more strenuous and best practices show $N_F/N_S=0.1$ to $0.5$ to be a more reliable sampling compression limit for larger datasets.
    \item \textbf{POD-SFR number of POD modes ($r_F/N_F$)}: Keeping the spatial and temporal sampling rates constant and using the POD-based ROM method of Secn.\ref{secn:method}, multiple orders of magnitude compression are achievable on top of the already down-sampled data. Here, only $30$ POD modes ($r/N_F=0.123$) accurately preserved rare event dynamics, although this $C=563.4$ level of compression is again dependent on the size and rank of the data.
    
\end{enumerate}

\section{Summary and Conclusions}\label{secn:conclusions}

Several sparse flow reconstruction (SFR) methods are presented to reduce the cost of writing, storing, and analyzing large unsteady datasets.
The fundamental concept of SFR is to write two smaller, sparse datasets, one with far fewer randomly placed sparse measurements, $\textbf{Y}$, written at the highest temporal sampling rate desired. 
The second is the full sized snapshots that are costly to write, $\textbf{X}$, and are therefore down-sampled in time $\textbf{X}'$. 
The general one-equation SFR model to approximate the full size snapshots at the high sampling rate is an interpolation projection of the sparse matrix onto a basis, $\mathbf{\tilde{X}=B(P^TB)^{-1}Y}$, where the basis consist of down-sampled full snapshots, $\textbf{B}=\textbf{X}'$, or proper orthogonal decomposition (POD) modes derived from these snapshots.
The overarching SFR method is presented as a post-processing tool that uncompresses (or reconstructs) the full solution for analysis.
Additional streaming algorithms were defined to improve the memory efficiency of parsing terabytes of data by exploiting the principles of sparsity and modal decomposition.


\subsection{SFR methods and features}
Two primary SFR methods are proposed: 1) the "snapshot method" and 2) a reduced order model of truncated POD modes, dubbed the "POD-SFR method". 
The snapshot SFR method is the simplest single-equation method where the available snapshots are the basis, while the POD-SFR method allows for greater data compression by truncating the basis to fewer POD modes. 
Two other SFR modifications can be implemented to improve the memory requirements of analyzing the full reconstruction on smaller computers.
The offline "streaming SFR" method breaks the SFR equation into multiple steps, producing a smaller SFR operator that can be stored in local memory, while individual snapshots are uncompressed one at a time as requested; this is necessary when the full CFD solution is larger than the available memory.
The second modification is specific to the POD-SFR method, and is considered a "double POD-SFR" calculation, where the burdensome decomposition of the POD modes from full snapshots is replaced with a smaller decomposition of the sparse measurements.
A subsequent SFR calculation reconstructs the truncated set of full-sized POD modes using the full snapshots, followed by a second SFR calculation that reconstructs the full snapshots from the truncated basis of POD modes. 
This variant can also be integrated with the streaming modification, eliminating the largest memory bottlenecks for post-processing data on the fly.

In addition to the fundamental SFR goal of reducing the data needed to analyze large unsteady solutions, several unique features and capabilities are noted to improve its implementation:
\begin{enumerate}
    \item Maximum accuracy at the highest sampling rate is obtained at sparse probe locations, $\textbf{Y}$. Therefore the sparse data should be substituted back into the approximated reconstruction after the SFR calculation. The preserved accuracy at these locations motivates the intentional placement of probes in important regions of the flow, in addition to the randomly scattered sparse probes. Increasing local sparse measurement density is practical and inexpensive, and should therefore be used judiciously.
    \item A proxy for global and transient error can be estimated by comparing sparse input measurements with reconstructed sparse measurements. This gives the user a metric to evaluate accuracy when no "truth" snapshots are available from the full solution to formally measure error.
    \item Sampling full and sparse datasets at rates with smaller coinciding intervals, $f_S/f_F$, creates a more accurate interpolation problem, but potentially aliases the data by the full sampling rate $f_F$. Contrary to this, consecutively writing out the full input snapshots at the highest sampling rate, $f_F=f_S$, over a short initial "training period", creates an overall less accururate extrapolation problem, but eliminates the aliasing error of interval sampling. Either full snapshot sampling approach is viable for any SFR method.
    
\end{enumerate}

\subsection{SFR performance and best practices}

The SFR methods were exemplified using LES data of a high-speed (Mach 1.7), high-Reynolds number ($Re=2.0 \times 10^6$) inlet/isolator/diffuser flow-path, featuring complex shocks and turbulence.
The SFR was evaluated by its ability to preserve rare, high-amplitude pressure distortion events, that require high sampling rates of the full snapshots over a long duration of time  to generate sufficient realizations of the events; however this comes with large memory costs in terms of generating and analyzing the database.  
SFR performance was measured in terms of the CFD solver writing penalty reduction, overall data compression, global and temporal reconstruction accuracy, and local preservation of rare event dynamics. 
These performance metrics are functions of the ratio of full to sparse sampling snapshots in time, $N_F/N_S$, the number of sparse measurements relative to the full snapshot size, $M_S/M_F$, and for the POD-SFR method, the number of POD modes used compared to the full snapshots available $r/N_F$.
Evaluation of these SFR parameters produced the following results and best practices to optimally balance compression and accuracy:

\begin{enumerate}
    \item The writing penalty improvement of SFR scales with decreasing the number of full snapshots sampled, $N_F/N_S$. The ambitious, $N_F/N_S=0.05$, $C=20$ compression rate for the inlet example reduced the worst-case writing penalty of the GPU-based LES solver from  $91\%$ to $32\%$. For the CPU-based LES solver, the less severe $29\%$ slowdown was effectively eliminated using the SFR method, enabling more efficient CFD.
    \item The primary source of error comes from decreasing the full snapshot sampling relative to the sparse sampling rates, $N_F/N_S$ and $f_F/f_S$. The minimum interval downsampling of every other full snapshot ($N_F/N_S=0.5$) produced an acceptable accuracy of $A=94\%$. In the extreme compression limit of using only three snapshots ($N_F/N_S=0.0002$), a stable accuracy of $77\%$ is maintained, but local turbulent structures associated with rare events are severely diminished.
    \item The fewer random and prescribed sparse measurements relative to the size of the full snapshot domain, $M_S/M_F$, contributes a much smaller source of local error than the temporal sampling rates if sufficient points are used. Results suggest a $1\%$ guideline of sparse measurements ($M_S/M_F>0.01$) to ensure this error is negligible, below which accuracy degrades. Placement of optimal sparse measurements over random points could further reduce this, but the $1\%$ allocation to random measurements is still favorably cheap.
    \item For the POD-SFR method, truncating the number of POD modes by half, $r/N_F=0.5$, doubles the SFR compression with minimal impact on error. Further truncation, can provide extreme compression for streaming snapshots on the fly, at the cost of only resolving the largest flow structures.
\end{enumerate}

Finally, note the exact accuracy and performance of the SFR results are specific to the case study presented, and results change according factors such as, the CPU or GPU hardware, the CFD solver, the size and rank of the database, and the complexity of flow physics.
However, the SFR trends demonstrated are likely universal and have shown a definitive ability to throttle between perfectly preserving space-time accuracy and orders of magnitude compression.
Further algorithmic and mathematical development of the fundamental SFR principles offers many opportunities for better data compression, speed, and memory improvements when analyzing large unsteady datasets.

\section*{Acknowledgments}\label{secn:ack}
This work was supported in part by high-performance computer time and resources from the DoD High Performance Computing Modernization Program. The authors would also like to acknowledge support and helpful conversations with D. Gaitonde.
S. Stahl was supported in part by an appointment to the NRC Research Associateship Program at the Air Force Research Laboratory, administered by the Fellowships Office of the National Academies of Sciences, Engineering, and Medicine. S. Benton acknowledges research funding from AFRL's Early Career Award used to fund software licenses.




\bibliography{./master.bib}

\end{document}